\documentclass[aip,jap,reprint]{revtex4-2}

\usepackage[english]{babel}
\usepackage[utf8]{inputenc}
\usepackage[top=2cm, bottom=2cm, left=2cm, right=2cm]{geometry}
\usepackage{color,soul}
\usepackage{graphicx}

\usepackage[svgnames]{xcolor}
\usepackage{textgreek}
\usepackage{array}
\usepackage{multirow}

\begin{document}


\title{Electron-stimulated desorption from molecular ices in the sub-keV regime}


\author{R. Dupuy}
\email{dupuy@fhi-berlin.mpg.de}
\affiliation{Sorbonne Universit\'e, Observatoire de Paris, Universit\'e PSL, CNRS, LERMA, F-75005 Paris, France}
\author{M. Haubner}
\affiliation{Dpt. of Physics, Faculty of Mechanical Engineering, Czech Technical University, Technicka 4, 16607 Prague, Czech Republic}
\affiliation{CERN, CH-1211 Geneva 23, Switzerland}
\author{B. Henrist}
\affiliation{CERN, CH-1211 Geneva 23, Switzerland}
\author{J.-H. Fillion}
\affiliation{Sorbonne Universit\'e, Observatoire de Paris, Universit\'e PSL, CNRS, LERMA, F-75005 Paris, France}
\author{V. Baglin}
\affiliation{CERN, CH-1211 Geneva 23, Switzerland}



\begin{abstract}
Electron-stimulated desorption (ESD) of cryosorbed molecules on surfaces is a process of relevance to fields as varied as vacuum dynamics in accelerators and astrochemistry. While desorption from such molecular systems induced by keV electrons and fast ions has been extensively studied, the sub-keV electron regime is comparatively little known. We measured and quantified electron-stimulated desorption from molecular ice systems (layers of N$_2$, CO, CO$_2$, Ar and H$_2$O/D$_2$O condensed at cryogenic temperatures) in the 150-2000 eV electron energy range. In this regime stopping power is no longer sufficient to explain the electron energy dependence of ESD yields. We introduce the notion of desorption-relevant depth, which characterizes the transition between two energy deposition regimes near the surface. We then apply this notion to the different systems, showing how ESD in the sub-keV regime can for example reveal the differences in species diffusion in crystalline and porous amorphous CO$_2$ ices.
\end{abstract}


\maketitle

\section{Introduction} 


Electron-stimulated desorption (ESD) is one of the oldest phenomena observed in vacuum science and has long been studied in the view of mitigating its impact in the deterioration of vacuum \cite{redhead1997,dempster1918}. Later more fundamental surface science studies of ESD were made to determine how the electronic excitations resulting from inelastic scattering of the incident electrons could convert to desorption, within the more general field of study of desorption induced by electronic transitions (DIET) \cite{menzel1995a,avouris1989,madey1994}. 

Studies of DIET phenomena still have important applications today in vacuum dynamics, in accelerators where synchrotron radiation creates energetic photons and electrons hitting the chamber walls \cite{grobner1999,hilleret2007}. Electron-stimulated desorption in particular is amplified by electron cloud effects\cite{cimino2014a}. In accelerators operated at cryogenic temperature, such as the LHC (Large Hadron Collider), the chamber walls are covered by molecules cryosorbed from the residual gas. These adsorbed molecules should change significantly the non-thermal desorption properties of the chamber walls. Some experiments were made previously on photon-stimulated desorption \cite{baglin2002} and electron-stimulated desorption \cite{tratnik2007} from cryosorbed gases with the aim to provide data to properly model vacuum behaviour in cryogenic instruments.

DIET phenomena also find applications in seemingly unrelated fields like astrochemistry. In cold and dense regions of the interstellar medium, where stars and planets are born, molecules condense from the gas phase or directly form on submicrometer-sized dust grains. These molecules and grains are exposed to irradiation from cosmic rays and stellar UV and X-rays, as well as from secondary photons and electrons of these primary sources. Non-thermal desorption induced by these various sources of energy regulate the exchanges between the gas phase and the solid phase at the surface of grains (see e.g. refs. \onlinecite{feraud2019a,dartois2018,brown1984,munozcaro2018a,dupuy2018c} and references therein). 

Experimental measurements providing absolute yields, but also a better fundamental understanding of non-thermal desorption processes, are necessary to the contexts where DIET phenomena are of relevance. The present study focuses on electron-stimulated desorption from molecular ices, i.e. on the high coverage regime where multiple layers of molecules are cryosorbed. We provide quantified data in the 150-2000 eV range, but we also aim to obtain more insights into some fundamental aspects of ESD, most notably to explain the electron energy dependence of desorption yields. The systems studied include molecules commonly found both in the residual gas of vacuum systems and at the surface of grains in the interstellar medium or other icy astrophysical objects (CO, CO$_2$, H$_2$O), as well as simpler molecular/atomic ices (N$_2$, Ar) where the absence of irradiation-induced chemistry allows to study these more fundamental aspects of ESD. 

Measurements of desorption by fast charged particles are usually rationalized by assuming that the desorption yield is proportional to some power of the stopping power for the particle-target couple considered. The stopping power is a measure of the energy loss per unit distance of a fast charged particle going through a given target material. It depends on the mass, velocity and charge of the particle as well as on the properties of the material. Many experimental studies have been performed to determine the relation between desorption yields and stopping power for molecular ices (see e.g. refs. \onlinecite{johnson2013,ellegaard1986,rothard2017,boduch2015,dartois2015a} and references therein). Most of these have been conducted with ions as projectiles, and sometimes with keV electrons. The electron energy dependence in the sub-keV regime has been comparatively little explored. Two studies to our knowledge report desorption yield curves as a function of electron energies in this range\cite{tratnik2007,huang2020}. Ref. \onlinecite{tratnik2007} does not expand on the physical interpretation of the observed dependence. In ref. \onlinecite{huang2020}, the energy deposition profile of electrons in the ice is described and used to explain the observed electron energy dependence of electron-induced chemistry in CO ice. The results on electron-stimulated desorption are however not discussed. Here it is our main goal to show how accounting for the electron energy dependence of ESD requires both the description of energy deposition by electrons as a function of energy, but also the energy and/or particle transport mechanisms that set which layers of the ice can be involved in desorption. Such a description therefore requires going beyond simple considerations of stopping power.

\section{Methods}

\subsection{Experimental set-up} 

\begin{figure*}
    \includegraphics[trim={1cm 2cm 9cm 0cm},clip,width=0.7\linewidth]{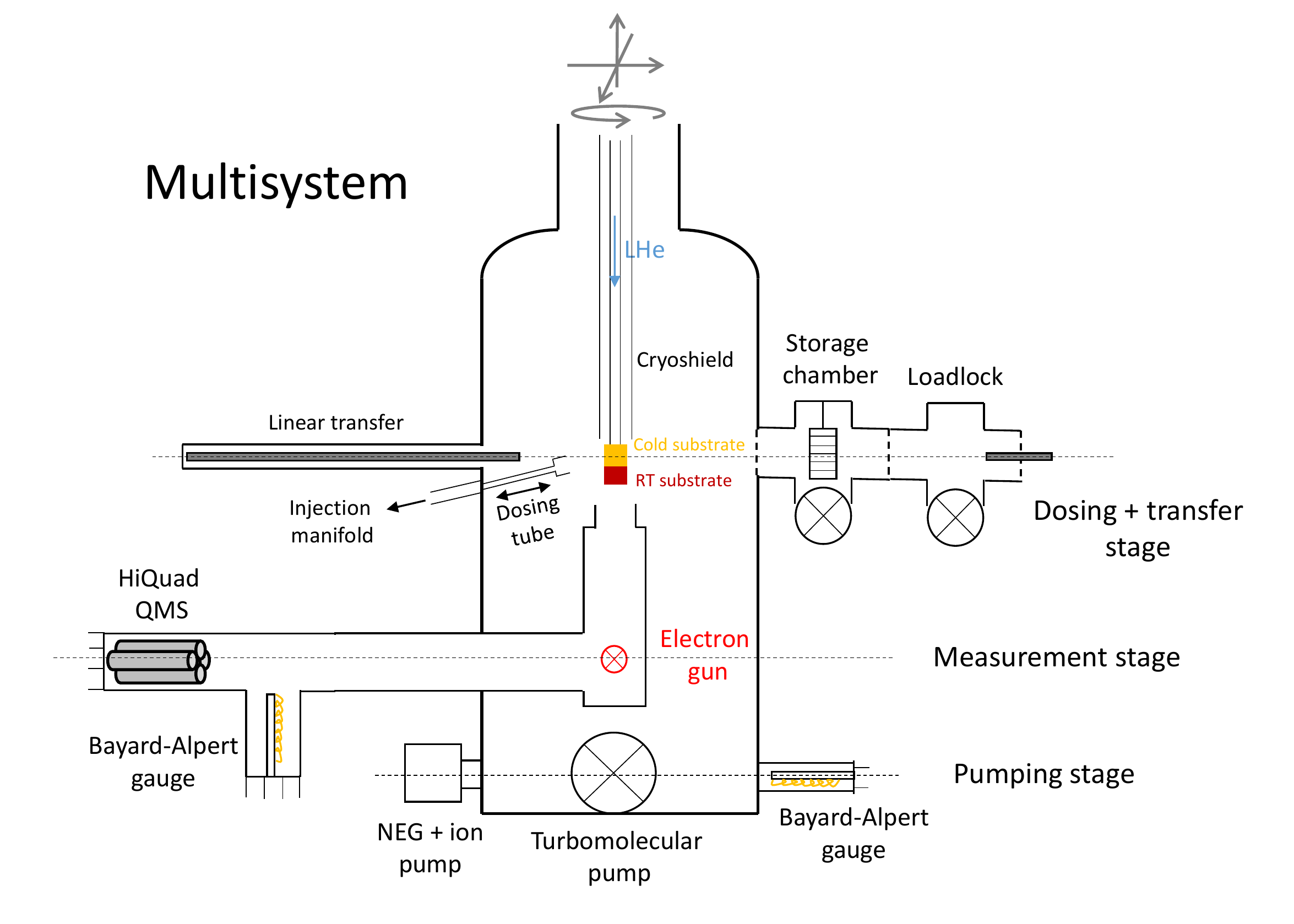}
    \caption{\label{Multisystem} Schematic drawing of the experimental Multisystem set-up. A description of the different elements is given in the text. QMS: Quadrupole Mass Spectrometer.}  
\end{figure*}

The Multisystem is a newly installed instrument at CERN. The main chamber of the system is an ultra-high vacuum mu-metal chamber pumped by a turbomolecular pump and a NEG (Non-Evaporable Getter) cartridge. The base pressure achieved is 1 $\times$ 10$^{-10}$ mbar at room temperature and 2 $\times$ 10$^{-11}$ mbar at 13 K. A drawing of the chamber is shown in fig. \ref{Multisystem}. At the center is a cryomanipulator with movements in the XYZ directions and rotation around the Z axis, cooled down with liquid helium. At the tip is a sample holder the upper part of which can be cooled down to about 13 K. Flag-type samples (surface $\sim$ 2.7 cm$^2$) can be inserted in the sample holder. For the experiments presented here the nature of the substrate sample is not considered important and its influence was not investigated. We used a gold substrate for the experiments ($>$ 1 $\mu$m gold vacuum-evaporated on a Cr covered stainless steel plate). The samples are kept electrically insulated by a sapphire plate, which allows to measure the currents produced by electron irradiation of, and electron emission from, the surface.

The chamber is equipped with an electron gun (Kimball), with a theoretical electron energy range of 5 - 2000 eV. The measurements were carried out in the 150 - 2000 eV range. The beam was incident normal to the surface. A phosphorus screen in the lower part of the sample holder and two Faraday cups on the back were used to characterize the electron beam. For each energy, the beam was set in order to obtain a clean spot of about 0.1 cm$^2$. The beam was checked to be gaussian with the Faraday cups. The exception was that for energies 1500 eV and more, the beam could no longer be focused tightly enough to obtain the same conditions. This resulted in larger and less well defined spots, which means measurements at these energies carry more uncertainty. 

The composition of the residual gas of the chamber was monitored using a quadrupole mass spectrometer (QMS, HiQuad), which is also used for the desorption measurements. 

In order to deposit condensed gas on the surface, an injection system is used. Gas from an injection manifold is brought into a closed, known volume where the pressure is measured by a capacitance gauge. A microleak valve is then used to let gas into the chamber through a dosing tube, which can be brought very close ($\sim$0.2 mm) to the surface so that most of the injected gas sticks and the rest of the chamber is not contaminated. From the QMS signal observed during ice deposition and using the calibration derived below, we estimated the flux leaking out of the substrate region during deposition to be about 1\% of the injected gas flux. The base pressure rises barely above 10$^{-10}$ mbar when injecting several hundred monolayers of ice. The investigated gases are N$_2$, CO, Ar, CO$_2$, and H$_2$O, as well as some isotopes of these: $^{15}$N$_2$, $^{13}$CO$_2$ and D$_2$O. For H$_2$O and D$_2$O, the liquids in pyrex vials were degassed by several freeze-pump-thaw cycles and the vapour pressure above the liquid was let into the manifold for injection. 

\begin{figure}
    \includegraphics[trim={0cm 0cm 0cm 0cm},clip,width=\linewidth]{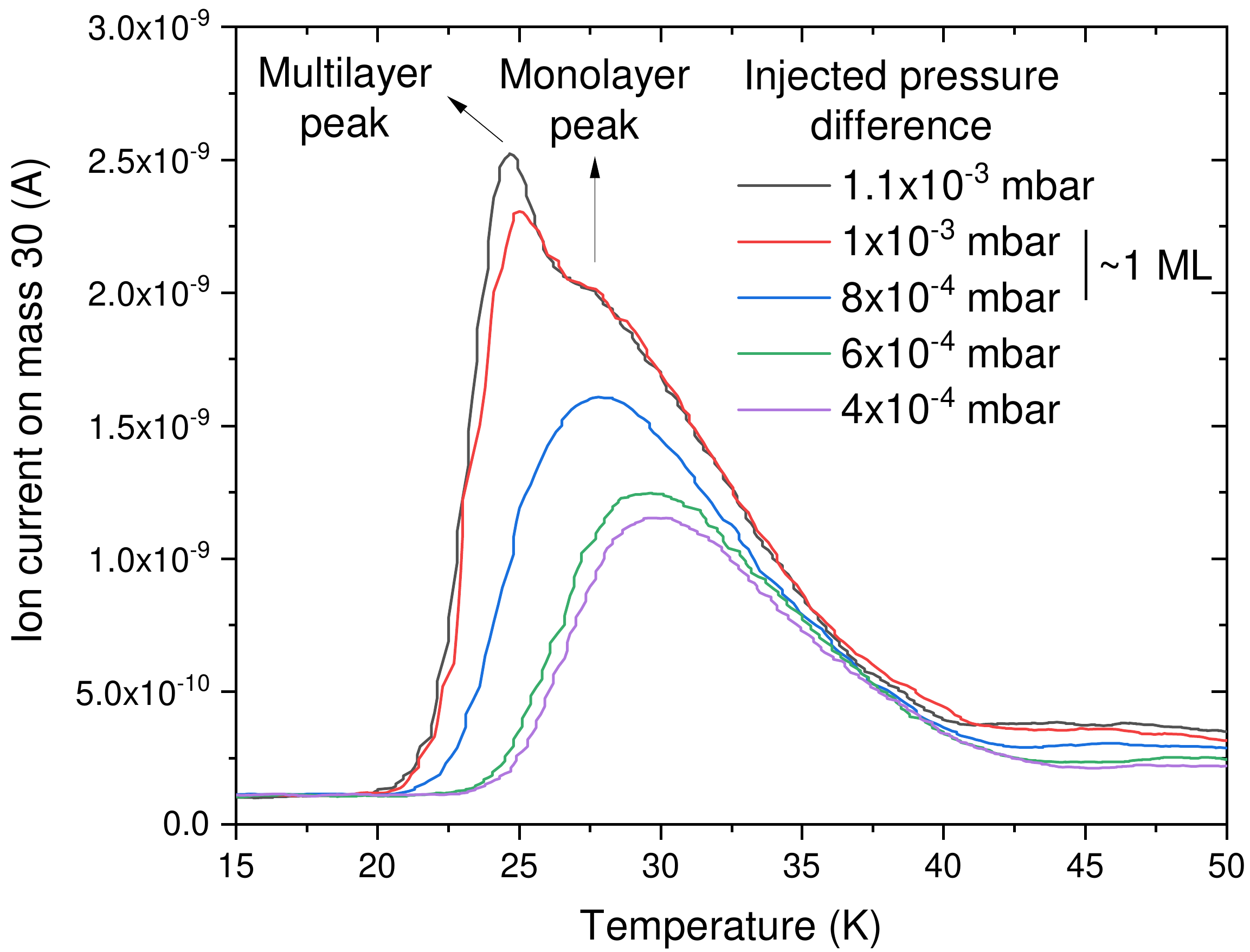}
    \caption{Calibration of the deposition of $^{15}$N$_2$ ice using TPD. A given amount of molecules, measured as a pressure drop in a known closed volume indicated in the legend, is deposited on the substrate, which is subsequently heated with a linear temperature ramp of 10 K/min, and the QMS signal is monitored as a function of temperature. The progressive appearance of a sharp peak at low temperature marks the onset of the multilayer regime.}
    \label{TPD_N2}  
\end{figure}

The amount of gas deposited is controlled by monitoring the drop in pressure from the closed calibrated volume. The thickness of ice corresponding to this deposited amount can be calibrated using the temperature-programmed desorption (TPD) method within $\sim$ 10-20 \% precision \cite{doronin2015}. The TPD of N$_2$ for different quantities of deposited gas on the substrate is shown in fig. \ref{TPD_N2}. For this case the transition from a sub-monolayer desorption regime to a multilayer regime is clearly identified by distinct desorption peaks, with a multilayer peak appearing at a temperature lower than the monolayer peak for the high coverages. This signifies that the desorption energy of the multilayer molecules is lower than the desorption energy of the monolayer molecules bound to the substrate. The pressure drop corresponding to the deposition of one monolayer can then be inferred from these TPD curves to be approximately 9 $\times$ 10$^{-4}$ mbar. One can also notice the desorption signal above 40 K in fig \ref{TPD_N2}, which is due to desorption from other cold parts of the manipulator that are progressively heated. This parasitic desorption does not contribute to the actual TPD signal from the surface. The thickness of ices for which the TPD method does not allow such a clear identification of the monolayer to multilayer transition (e.g. H$_2$O) was estimated from the N$_2$ calibration, taking into account differences in pumping speed.

One particularity of the set-up is the use of a "molecule collector", designed in the spirit of the so-called "Feulner cap" \cite{feulner1980}, to increase the sensitivity of the measurements. The measurement of electron-stimulated desorption involves measuring the increase of partial pressure of the species of interest when bombarding the surface with the electron beam. The collector is an L shaped tube attached to the flange of the chamber which leads to the QMS and into which the manipulator can be inserted (see fig. \ref{Multisystem}). The pumping speed inside the collector is thus limited, with the only apertures being the margin between the manipulator and the collector entrance, and a small hole to allow the electron beam to hit the sample inside. With the pumping speed reduced, the increase of partial pressure due to ESD when irradiating the sample is increased. Based on measurements made inside and outside the collector for N$_2$ ESD, it is estimated that the collector brings an improvement of a factor of $\sim$ 8 to the signal-to-noise ratio of the QMS signal.

\subsection{Measurement procedure}

We wish to obtain the ESD yield, i.e. the number of ejected molecules or atoms per incident electron, as a function of electron energy for a number of systems. For this, a molecular ice is grown by the procedure explained previously. The cryomanipulator is then inserted in the molecule collector so that the cold substrate is in front of the electron gun. The surface is set at a bias voltage of +45 V to prevent secondary electrons from escaping the surface. The current read on the substrate when irradiating is therefore the incident electron current. The bias voltage had no influence on the ESD yields other than that due to the change of incident electron kinetic energy, which is taken into account. 

The ice is irradiated at a given electron energy for about 20-30 seconds, usually with a beam current of around 100 nA (flux of electron $\sim$ 6 $\times$ 10$^{11}$ e$^-$.s$^{-1}$), and the desorption signal of the relevant species is recorded on the QMS as an increase of partial pressure. Simultaneously, the substrate (incident beam) current is recorded with a Keithley picoammeter. The procedure is repeated on the same ice spot (since the molecular collector does not allow to move the sample) for a series of electron energies. We checked that the order in which points were measured did not influence the desorption yields measured.

Once the data is recorded, the average value of the QMS current i$_{QMS}$ during irradiation for each energy is divided by the beam current \textPhi$_{e^-}$, then converted to an absolute ESD yield Y:

\begin{equation}
Y_X = \frac{i_{QMS}}{k_{X}\Phi_{e^-}}
\end{equation}

The calibration of QMS current to ESD yield to derive the calibration factor k$_{X}$ for species X is made using the temperature-programmed desorption (TPD) technique. This calibration technique is described e.g. in ref. \onlinecite{dupuy2018c}. It consists in assuming that the calibration factor k$_{X}$ between QMS current and desorption flux is also valid in a thermal desorption experiment such as TPD. Since the ice thickness deposition is calibrated (using TPD) we can deposit a known absolute number of molecules on the substrate and evaporate it in a TPD experiment. Then we have:

\begin{equation}
 k_X = \frac{\int i_{TPD}dt}{N_X^{1 ML}}
\end{equation}

where N$_X^{1 ML}$ is the total number of molecules for 1 monolayer (ML = $\sim$ 10$^{15}$ mol.cm$^{-2}$ $\times$ 2 cm$^{2}$ for all the gases considered here) and i$_{TPD}$ is the signal read on the QMS during the TPD experiment where exactly 1 ML is sublimated. For example, for N$_2$ we derived a value of k$_{N_2}$ = 1.05 $\times$ 10$^{-23}$ A.s. For species for which a TPD experiment is not possible (e.g. H$_2$), the calibration factor can be extrapolated from the one of another molecule by taking into account the relative sensitivity of the QMS to the two different species. This relative sensitivity is measured by injecting known pressures (using a calibrated gauge) of the respective gases in the chamber and measuring the corresponding QMS signals.

The root-mean-squared noise of the QMS measurement is taken as a measurement uncertainty for the error bars in the subsequent figures. We thus consider that there is no uncertainty on the beam current measurement. This assumption can be wrong, in particular if there is reflection of electrons on the gold substrate and subsequent escape from the ice, which would lead to an underestimation of the true incident current. This point is also discussed later in the text. As for the absolute yield calibration, we estimate a $\sim$ 50 \% uncertainty.  

\section{Results}

Electron-stimulated desorption was studied and quantified between 150 and 2000 eV for five molecular ice systems: N$_2$, CO, Ar, CO$_2$ and H$_2$O (or D$_2$O). The results for each of these systems will first be presented separately. Further interpretation of these results, in particular of the electron energy dependence, in the frame of the notion of desorption-relevant depth, will be made in the Discussion part. 

\subsection{N$_2$}

\begin{figure}
	\centering
    \includegraphics[trim={0cm 0cm 0cm 0cm},clip,width=\linewidth]{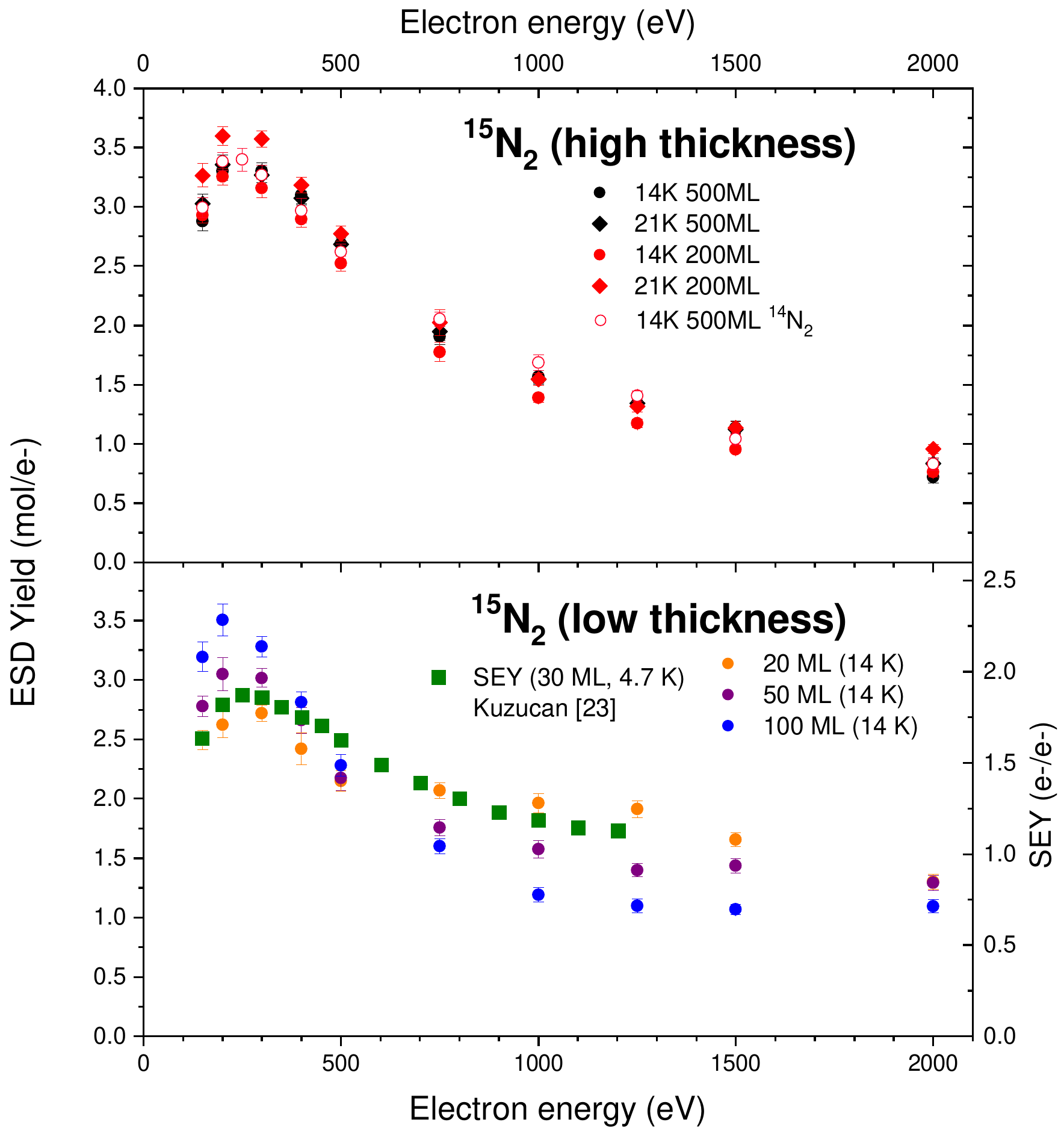}
    \caption{$^{15}$N$_2$ ESD yield for different ice thicknesses and deposition temperatures. Upper panel: curves for 500 ML (monolayer) and 200 ML ices and for deposition temperatures of 14 and 21 K, as well as a curve measured on $^{14}$N$_2$. Lower panel: curves for 20, 50 and 100 ML ices deposited at 14 K. The secondary electron yield (SEY) of 30 ML N$_2$ ice from ref. \onlinecite{kuzucan2012} is also shown.} 
    \label{N2_EID_thickness}
\end{figure}

Experiments were performed on both regular N$_2$ and isotopically labeled $^{15}$N$_2$. No differences were observed, and only the results for $^{15}$N$_2$ will be presented here, except for one curve measured on $^{14}$N$_2$ to substantiate the absence of difference between the isotopes. The only desorbing species observed was N$_2$ itself, although it is possible that very small amounts of N, N$_3$ or N$_4$ desorb as well \cite{pedrys1989}. Nitrogen ice exhibits green luminescence when irradiated with electrons, which allows to cross-check the shape and position of the electron spot directly on the sample. 

On the upper panel of fig. \ref{N2_EID_thickness}, results for four different ices are shown, for 200 and 500 ML (monolayer) thicknesses and for a deposition temperature of the ice of 14 and 21 K. All the curves are very similar. Therefore the deposition temperature of the ice does not have an effect here, and the ESD yields already correspond to the semi-infinite ice values at 200 ML. We also checked (not shown) that the temperature at which the ice is maintained during the irradiation does not have an effect either (from 14 to 22 K: above 22 K thermal desorption becomes higher than ESD). The only identified structural change for N$_2$ ice occurs around 35 K \cite{vegard1929}, a temperature outside the range that can be accessed under high vacuum conditions. There are no other indications of structural changes in the range of deposition temperatures explored here (for example the density remains constant \cite{satorre2008}), therefore these results are not surprising. The curves exhibit a maximum around 300 eV.

Results for thinner ices are shown on the bottom panel of fig. \ref{N2_EID_thickness}. When we reach down to 20 ML, the curve "flattens" with a higher yield at high electron energies ($>$ 500 eV) and a slightly lower yield around the maximum ($<$ 500 eV). To explain why the curve flattens for thinner films, we need to take into account the substrate. Indeed, electrons with a penetration depth higher than the film thickness will reach the substrate. Excitations of the substrate could then take place and influence the yield, but another factor is the reflection of electrons on the substrate. In particular, gold has a high specular reflection coefficient for electrons, e.g about 0.4 at 2 keV \cite{so/rensen1978}. Two distinct consequences of electron reflection can lead to the observed increase (or apparent increase) of the desorption yield at high electron energy, if the penetration depth of the electrons is at least twice the thickness of the film (i.e. electrons can be reflected and reach the surface again). The first is that electrons will then re-deposit energy close to the surface, leading to a higher yield than in the semi-infinite ice case. The second is that if reflected electrons escape the surface, this will lead to an underestimation of the true incident current, and thus to an apparent increase of the ESD yields.

On the figure is also shown a measurement of the secondary electron yield (SEY) of N$_2$, taken from Kuzucan et al. \cite{kuzucan2012}. The SEY curve, which was measured for a 30 ML ice at 4.7 K, closely resembles the ESD curves of about the same thickness (20 ML), suggesting similar processes drive ESD and SEY in this case.   

The absolute ESD yield for N$_2$ reaches about 3.3 molecules/electron at the maximum. The yield at 1 keV is 1.6 molecules/electron, which compares very well with the value of 1.8 molecules/electron found in ref. \cite{ellegaard1986} and quantified using a quartz microbalance. In ref. \cite{tratnik2007} the value indicated is about 2.5 molecules/electron, which is also close. There are other studies of ESD from N$_2$ \cite{pedrys1989,hudel1992,savchenko2017a,savchenko2017} which however did not report quantitative estimates, focusing instead on other phenomena such as desorption of excited species, kinetic energy distribution measurements or luminescence.  

\subsection{CO}

\begin{figure}
	\centering
    \includegraphics[trim={0cm 0cm 0cm 0cm},clip,width=\linewidth]{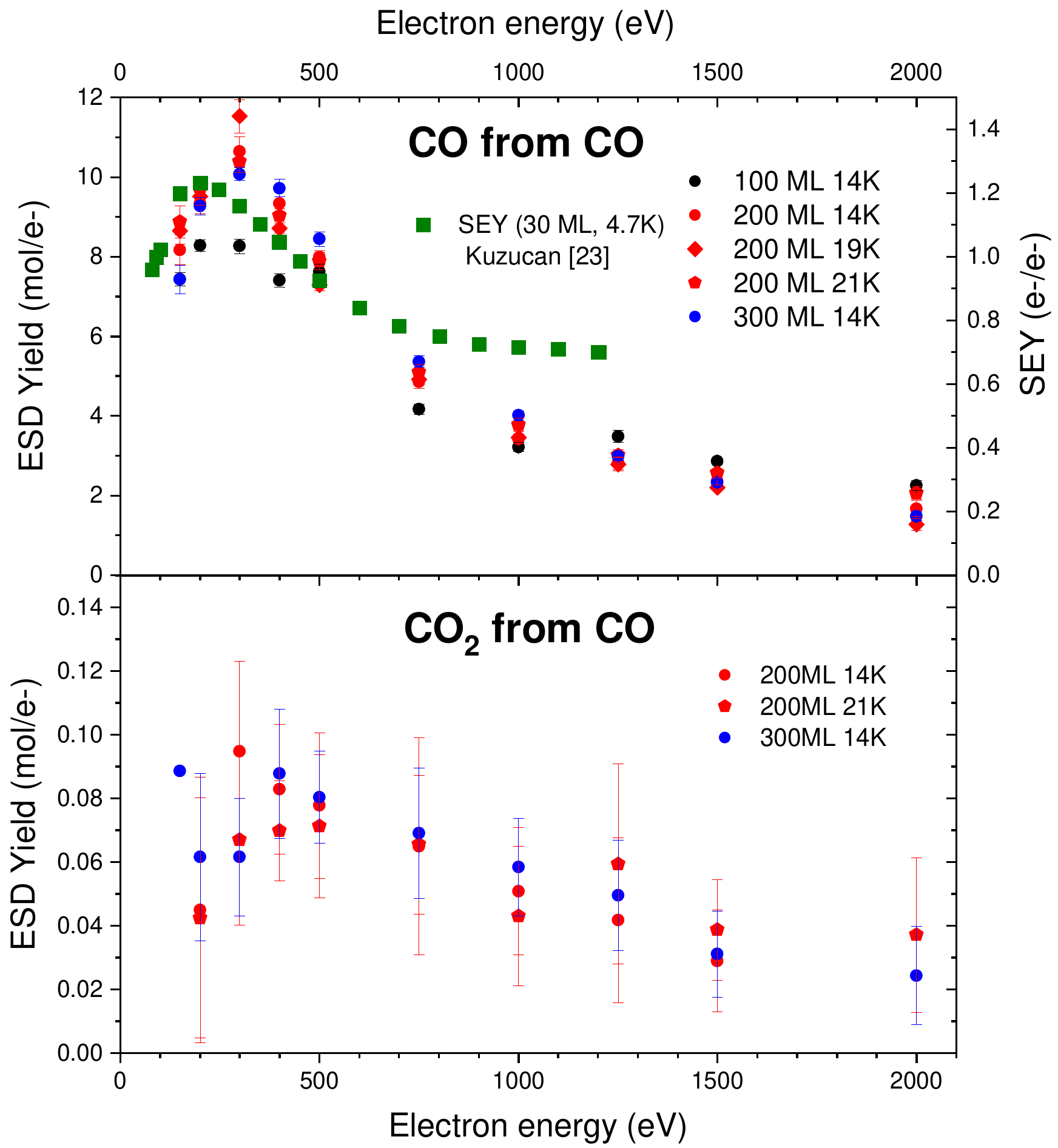}
    \caption{Upper panel: CO ESD yield from CO ice for different thicknesses and deposition temperatures. The SEY of 15 ML CO from ref. \cite{kuzucan2012} is also shown. Lower panel: CO$_2$ ESD yield from CO ice for different thicknesses and deposition temperatures.} 
    \label{CO_EID}
\end{figure}

The ESD yield for CO ices of different thicknesses and deposition temperatures are shown in fig. \ref{CO_EID} (upper panel for CO desorption). Once again, a typical curve with a maximum at 300 eV is observed. The yield, however, is much higher than for N$_2$. The yields have already reached a semi-infinite ice value at 100 ML. Measurements of CO ESD for thinner films were made difficult by the high ESD yield of the molecule and the modification of the film during irradiation: too much of the ice was processed or removed during the measurement (for an ESD yield of 10 mol/electron and an irradiation at 100 nA on a 0.1 cm$^2$ spot the removal rate is about 3 ML/minute). No effect of the deposition temperature (14-21 K) is observed. Also shown on fig.\ref{CO_EID}a for comparison is the CO SEY from Kuzucan et al. \cite{kuzucan2012}, which here differs slightly from the ESD curves, both at low and high energies. This SEY curve was measured at a thickness of 15 ML, which explains the discrepancy at high electron energies ($>$ 500 eV), as specular reflection of electrons and substrate contribution will play a role here as well. At low electron energies, the maximum of the SEY occurs at 200 eV and is therefore different from the 300 eV maximum of the ESD, which will be explained later. Our value of $\sim$3.8 CO molecules/electron at 1 keV is close to the value of Tratnik et al. \cite{tratnik2007} ($\sim$4.5 molecules/el) but higher than the values reported by Schou et al. and Brown et al. \cite{brown1984} which are below 1 molecule/electron.  

Contrary to N$_2$ ice, a lot of chemistry can already happen during the irradiation of CO ice, as many carbon chain species can be formed. Refs \cite{jamieson2006,huang2020} are examples of studies of the irradiation-induced chemistry in CO ice, which includes formation of many C$_n$, C$_n$O and C$_n$O$_2$ species. Although we cannot probe the chemistry happening in the ice here, we can observe one of the main products desorbing in the gas phase, CO$_2$. No other species are observed desorbing. The ESD yield of CO$_2$ from a CO ice is shown in the lower panel of fig. \ref{CO_EID}. We observed a progressive increase of CO$_2$ desorption over the course of a few minutes after the beginning of irradiation, due probably to the accumulation of CO$_2$ in the ice. Desorption reaches a constant value after some time and the yields shown correspond to that steady state value. The yield is very small (two orders of magnitude smaller than CO) and therefore the error bars are large, so it is difficult to confirm whether there is a difference in the shape of the ESD curves between CO and CO$_2$. 

\subsection{Ar}

\begin{figure}
	\centering
    \includegraphics[trim={0cm 0cm 0cm 0cm},clip,width=\linewidth]{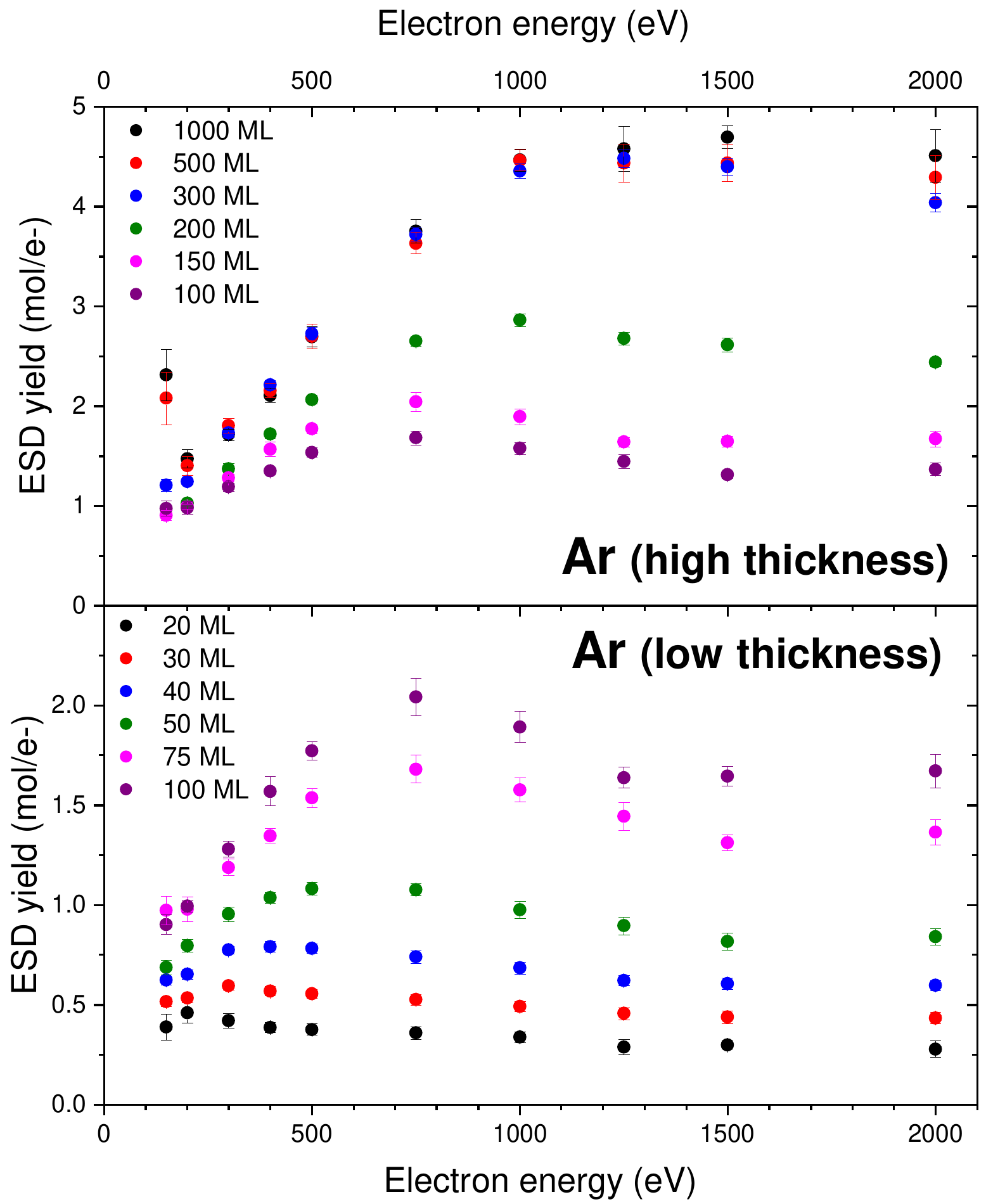}
    \caption{Ar ESD yield for different ice thicknesses. High thicknesses are shown on the upper panel and low thicknesses on the lower panel. All ices were deposited and irradiated at 14 K.} 
    \label{Ar_EID_thickness}
\end{figure}

The ESD yield of solid argon as a function of electron energy and ice thickness is displayed in fig. \ref{Ar_EID_thickness}, with the high thicknesses on the upper panel and the low thicknesses on the lower panel. The shape of the ESD yield curve changes depending on the thickness up to about 300 ML, with the maximum shifting to higher energies. Above 300 ML the shape and intensity become stable, with only marginal changes for the highest electron energies. The shape of the desorption curve of Ar ice at high thickness is markedly different from what is observed for the two previous ices.

Quantitatively, the yield of 4.5 Ar atoms/electron at 1 keV is in good agreement with previously reported results by Tratnik et al.\cite{tratnik2007} ($\sim$ 5 atoms/electron) and Ellegaard et al.\cite{ellegaard1988} (2.5 atoms/electron) for thick argon films. 

\subsection{CO$_2$}

\begin{figure}
	\centering
    \includegraphics[trim={0cm 0cm 0cm 0cm},clip,width=\linewidth]{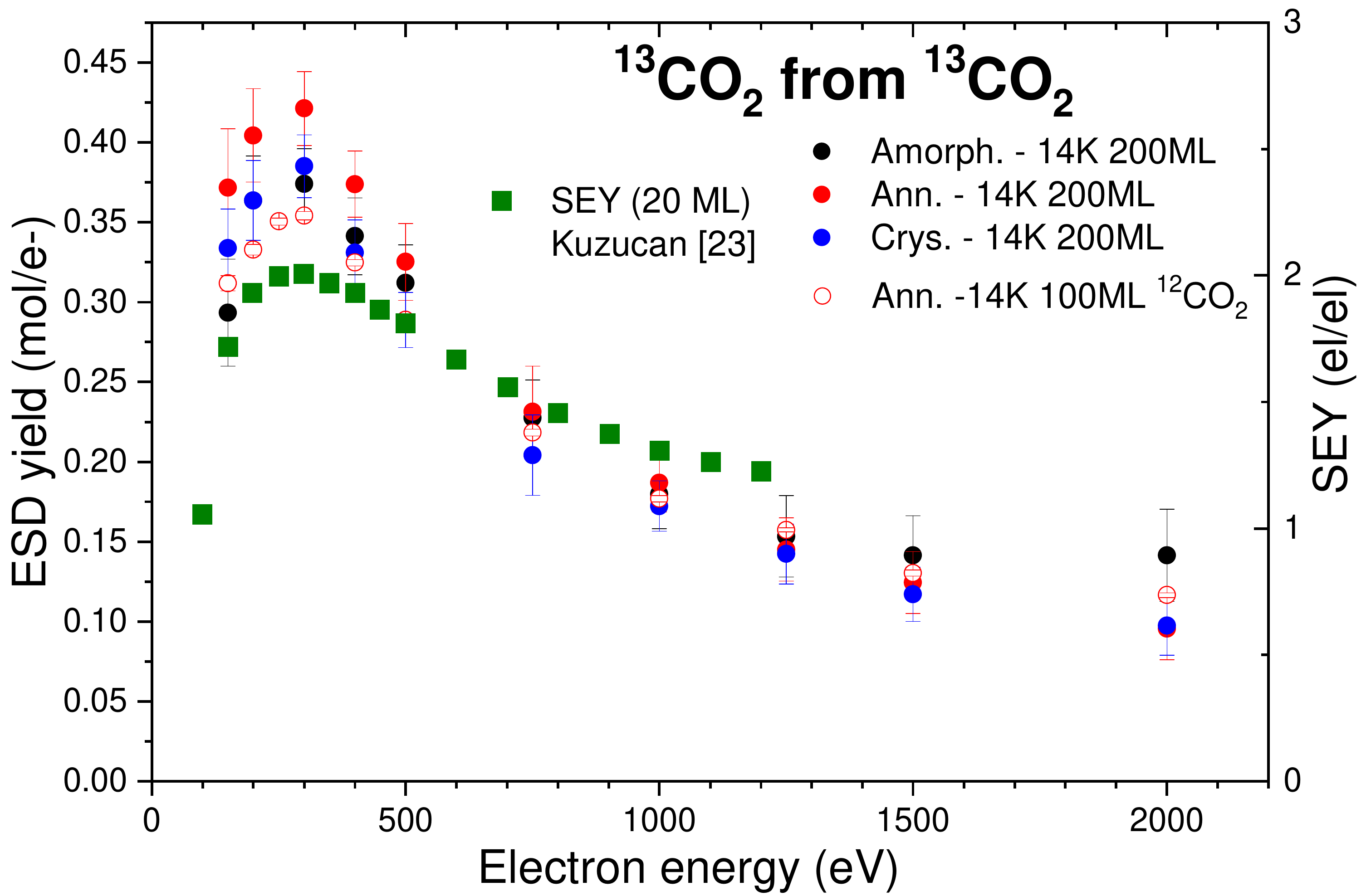}
    \caption{ESD yield of $^{13}$CO$_2$ from $^{13}$CO$_2$ ice, for three different structures of the ice (see the text). Also shown is a curve measured on $^{12}$CO$_2$. In the labels, the temperature indicates the irradiation temperature. Also shown is the SEY for CO$_2$ ice from ref. \cite{kuzucan2012}.}
    \label{CO2_EID}
\end{figure}

Solid carbon dioxide has a complex range of possible structures. CO$_2$ films can be crystalline or amorphous to various degrees depending on the conditions of preparation of the ice. The ice is amorphous when deposited at a temperature of 14 K (the base temperature of our set-up) and becomes gradually more and more crystalline for deposition temperatures up to 75 K or when it is annealed to such temperatures (above this temperature, under our experimental conditions, the ice starts to significantly sublimate) \cite{he2018}. Deposition temperature also significantly affects the density and porosity, with an ice 40\% less dense when deposited at 14 K than when deposited at 75 K \cite{schulze1980}. 

Experiments were done on both regular CO$_2$ and isotopically labeled $^{13}$CO$_2$. As for N$_2$ only the experiments with $^{13}$CO$_2$ will be presented here - except for one curve on $^{12}$CO$_2$ - as no differences were observed between the isotopes. Three different phases of the ice are investigated: one where the ice is grown at 14 K and irradiated as is (Amorph.), one where the ice is grown at 14 K then annealed for 5 minutes to 75 K (Ann.), and one where the ice is directly deposited at 75 K (Crys.). The latter two show crystalline properties but are different, which is seen from the fact that CO$_2$ ice directly deposited at 75 K can be seen with the naked eye on the surface. A 200 ML ice is usually too thin to be visible. Here a very clear and colored square with iridescence is observed. Presumably the ice grown at 75 K is the most crystalline one and the crystalline order is at the origin of this optical property. In the absence of more structural insights, we do not presume any of the properties of the annealed and crystalline ices, such as their degree of crystallinity (one or both could still be partially amorphous) or a difference in structure. It should be noted that in the case of the crystalline and annealed CO$_2$ ice, which are respectively grown at 75 K or annealed for some time at that temperature, the actual ice thickness is lower than the nominal thickness since the temperature is above the sublimation temperature of the ice. In the figures to follow, the ices are labeled Amorph., Ann. or Crys. according to the method of growth. All ices were irradiated at 14 K. 

The results for desorption of CO$_2$ are displayed in fig. \ref{CO2_EID}. The yields and shape of the curves for the three phases are very close. The curves are again compared with the SEY results of Kuzucan et al. \cite{kuzucan2012}, which have the same maximum at 300 eV. The SEY curve is flatter (higher at high electron energies relative to the maximum) but this is again ascribed to thickness differences, as the SEY was measured for a 20 ML ice. 

\begin{figure}
	\centering
    \includegraphics[trim={0cm 0cm 0cm 0cm},clip,width=\linewidth]{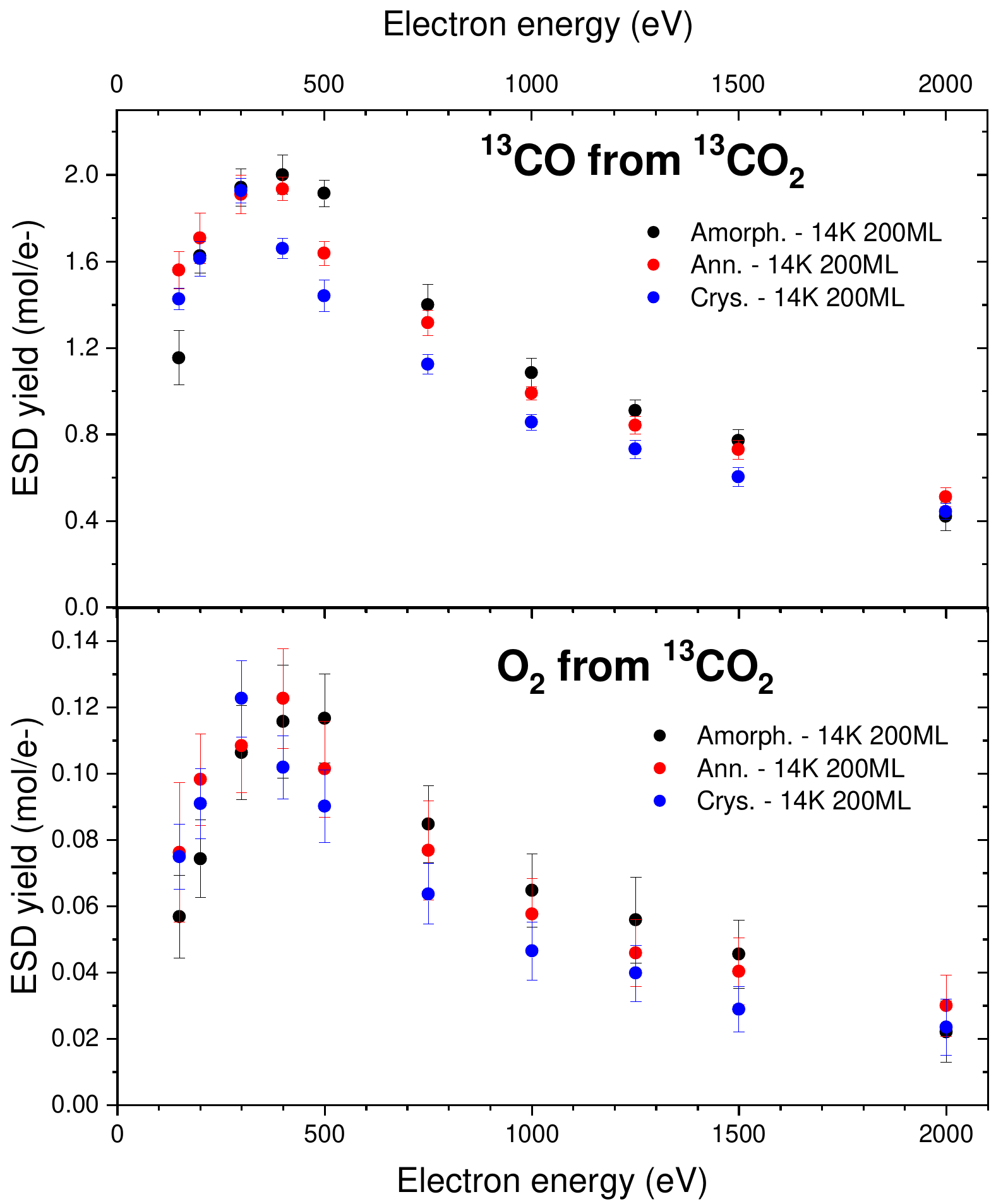}
    \caption{ESD yield of $^{13}$CO and O$_2$ from $^{13}$CO$_2$ ice for different structures of the ice. In the labels, the temperature indicates the irradiation temperature.} 
    \label{CO2_CO_O2_EID}
\end{figure}

The most abundant desorbing species is not the parent molecule CO$_2$ but a product of dissociation, CO. Also observed desorbing is O$_2$, although the yield is lower. This is consistent with other studies of desorption from CO$_2$, whether the ones with ions \cite{johnson2013} or the ones with VUV photons \cite{fillion2014,sie2019}. The desorption yields of CO and O$_2$ for the same ices that were presented in the previous figure are shown in fig. \ref{CO2_CO_O2_EID}. Similarly to the CO case, there is a time dependence for the desorption of these products with a gradual increase over a few minutes before reaching a plateau, and we display and discuss the plateau value here. Although the absolute yields do not differ significantly for different structures of the ice (which is consistent with the result of Sie et al. for VUV photons \cite{sie2019}), there is a difference in the shape of the curves. As we go from the more crystalline ice (deposited at 75 K) to the more porous and amorphous one (deposited at 14 K), there is a shift towards higher energies of the maximum (CO: 300 eV for Crys., between 300 and 400 eV for Ann, 400 eV for Amorph.; O$_2$: 300 eV for Crys., 400 eV for Ann., 500 eV for Amorph.).

\subsection{H$_2$O/D$_2$O}

\begin{figure}
	\centering
    \includegraphics[trim={0cm 0cm 0cm 0cm},clip,width=\linewidth]{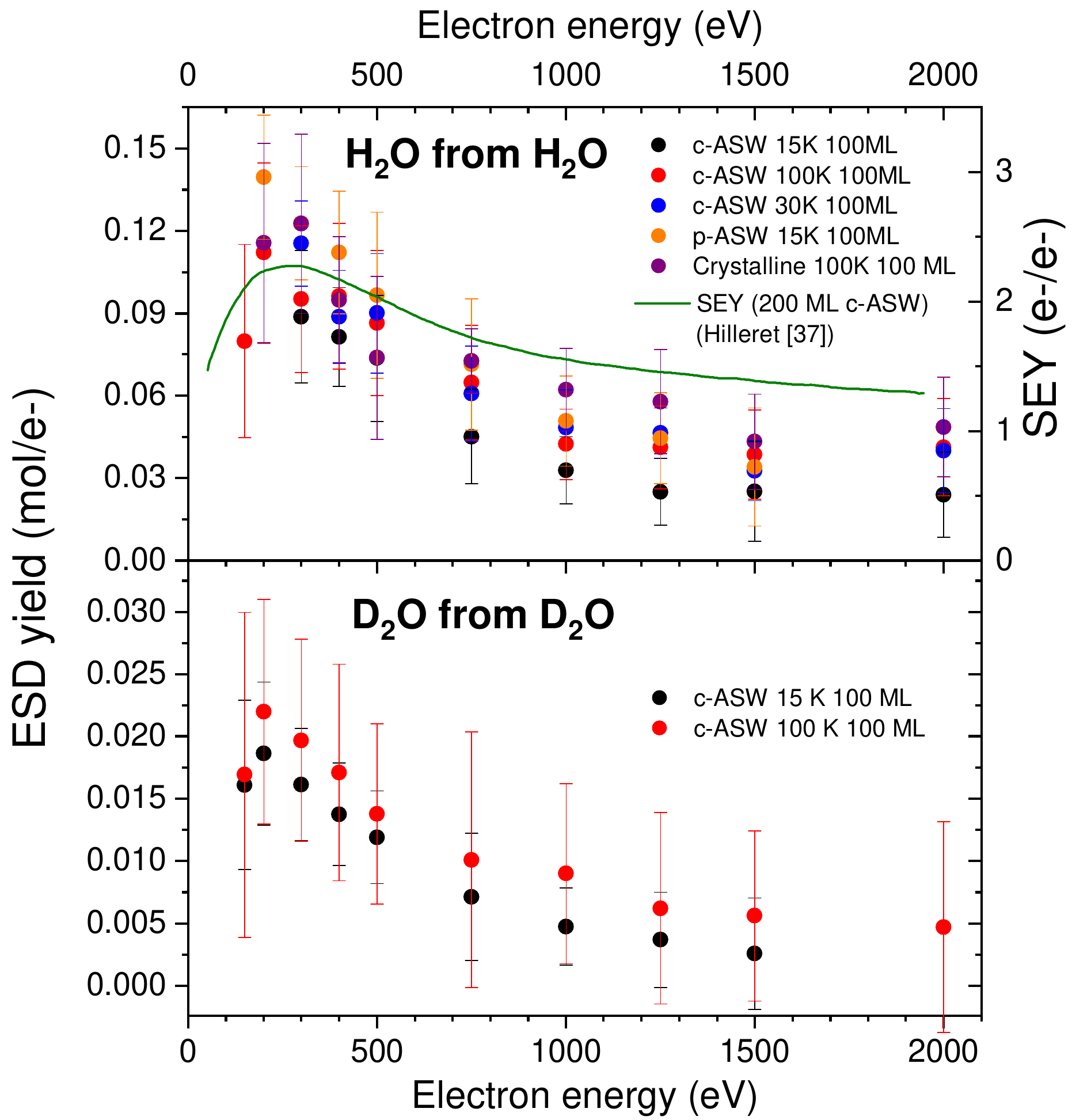}
    \caption{Top panel: ESD yield curve of H$_2$O from water ice, for different phases and irradiation temperatures. The SEY from ref. \cite{hilleret2000} is also shown. Bottom panel: ESD yield curve of D$_2$O from c-ASW D$_2$O ice, at two different irradiation temperatures.} 
    \label{H2O_D2O}
\end{figure}

The ESD yield curve of H$_2$O from water ice is shown in fig. \ref{H2O_D2O}a for different phases and temperatures of the ice. The ice is grown at 15 K to obtain porous amorphous solid water (p-ASW), at 100 K to obtain compact amorphous solid water (c-ASW) and 140 K for crystalline water. The irradiation temperature can then also be varied. It is clear from fig. \ref{H2O_D2O}a that the phase does not have significant effects on the desorption of H$_2$O. The yields for higher irradiation temperatures are systematically higher, but not by much. The SEY of water ice adapted from ref. \onlinecite{hilleret2000} is also shown on fig. \ref{H2O_D2O}a. While the maxima look similar between the SEY and ESD curves, the shape are not exactly the same but it is difficult to comment further, especially considering the error bars on the points. 

\begin{figure*}
	\centering
    \includegraphics[trim={0cm 0cm 0cm 0cm},clip,width=0.8\linewidth]{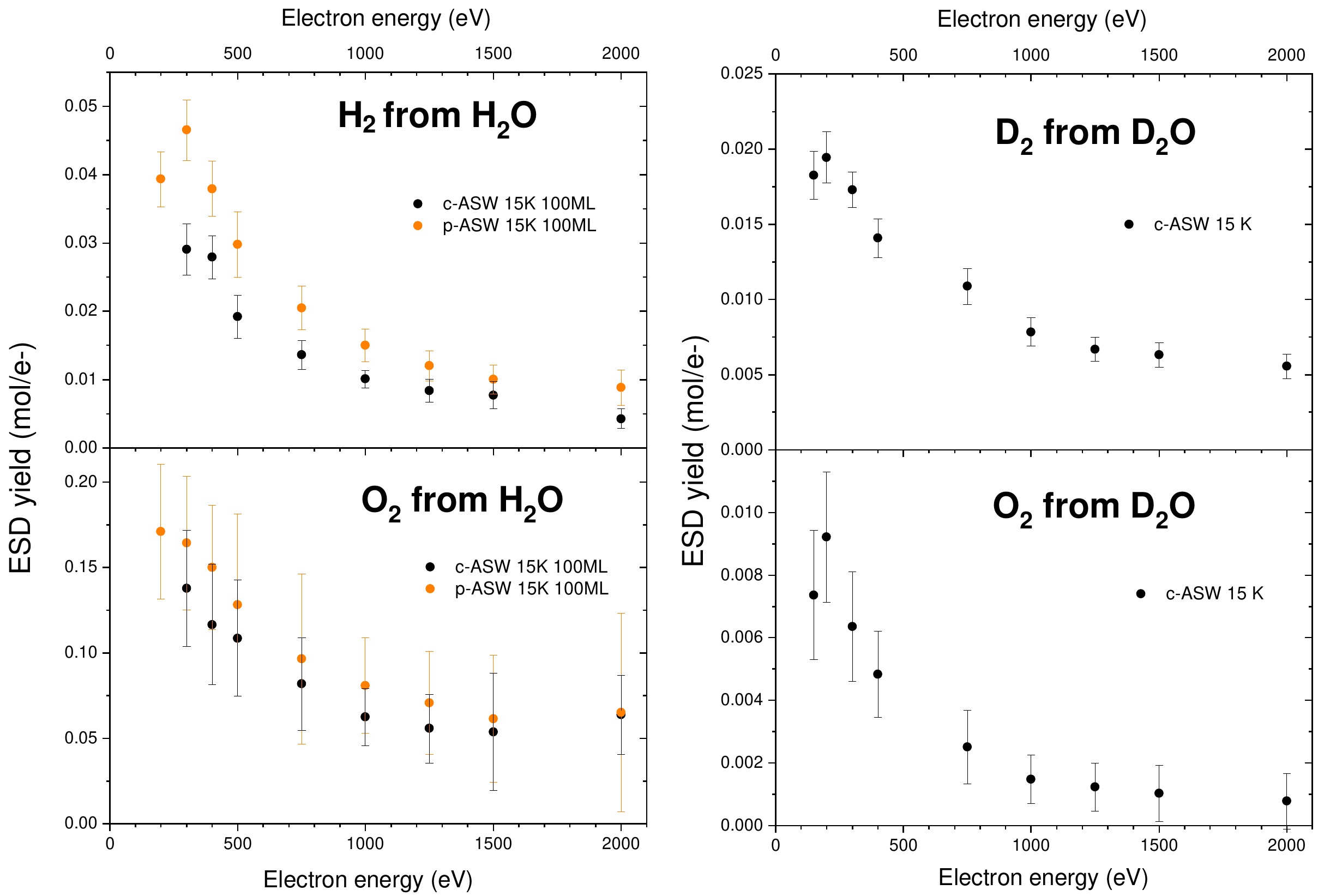}
    \caption{ESD yield curves of H$_2$ (top left) and O$_2$ (bottom left) from H$_2$O ice for c-ASW and p-ASW at 15 K, and D$_2$ (top right) and O$_2$ (bottom right) from D$_2$O ice for c-ASW at 15 K.} 
    \label{H2O_D2O_prod}
\end{figure*}

We have also made experiments on D$_2$O ice, and the results displayed in fig. \ref{H2O_D2O}b that show the ESD yield of D$_2$O from compact amorphous D$_2$O ice at 15 or 100 K show an isotope effect in the absolute yields. We find a factor of $\sim$ 5 between the desorption yields of H$_2$O and D$_2$O. A similar factor was found for VUV desorption of water \cite{dupuy2019e}, although the explanation for this observation is not clear at the moment.

On fig. \ref{H2O_D2O_prod}, the results obtained at 15 K for the desorption of H$_2$/D$_2$ and O$_2$ from H$_2$O/D$_2$O ices are shown. The desorption of H$_2$ and O$_2$ show time dependences as well, and again points were taken when the signal had reached a relatively stable value. The yield of H$_2$O, measured simultaneously, did not show any time dependence. Results for the desorption of H$_2$ and O$_2$ at higher temperatures are not shown because of quantification and time dependence issues. 

Considerable amount of work has been made on water ice ESD, to which the reader is referred for more details. To cite only those that consider neutral species desorption and without being exhaustive, there have been studies of ESD by low-energy ($<$ 20 eV) electrons \cite{kimmel1994,kimmel1995}, and numerous studies of ESD by sub-keV electrons in the group of Petrik and Kimmel \cite{petrik2004,petrik2005,petrik2006a} and more recently in the group of McCoustra \cite{marchione2016a,abdulgalil2017}, that point to the complexity of the energy transport and chemistry leading to production of H$_2$ and O$_2$, in particular.

Quantitative studies of water ESD\cite{petrik2005,abdulgalil2017} agree with our results at least in the order of magnitude, although these works look at the H$_2$O loss from the bulk and therefore constitute only an upper limit on the water yield. The total estimated loss of 0.5 molecules/electron in ref. \onlinecite{petrik2005} for 87 eV electrons incident on D$_2$O at low temperature is much higher than our D$_2$O yield at low electron energies, around 0.02 molecules/electron. The difference could be explained by loss of D$_2$O through desorption of products of chemistry. No absolute sputtering yields are reported for H$_2$O in this study. The number of 0.1-0.5 molecules/electron for 200-300 eV electrons on H$_2$O in ref. \onlinecite{abdulgalil2017} is close to our derived yield.

\section{Discussion}

\subsection{Desorption-relevant depth}

In the electronic stopping regime - which concerns electrons and fast ions - particles interact with the electrons of a material and deposit energy mainly in the form of electronic excitations. This interaction is characterized by the stopping power, which is the amount of energy deposited by the particle in the material per unit distance traveled. Desorption induced by particles in the electronic regime is therefore assumed to be related to the stopping power of the particle-target pair. The underlying assumption here is that the desorption yield depends in some way on the amount of energy deposited at or "near" the surface, since desorption is a surface effect, and this amount of energy is by definition proportional to the stopping power. Indeed, the observed desorption yields as a function of particle energy - and therefore of stopping power - usually verify a power law, the dependence often being linear or quadratic\cite{schou1987,johnson2013}. A linear relation can be interpreted in terms of domination of a mechanism of desorption involving independent individual events (event = excitation or ionization of a molecule), although such mechanisms do not necessarily lead to a strictly linear relation depending on the details of energy relaxation in the system considered. A superlinear (often but not necessarily quadratic) relation instead suggests mechanisms of desorption involving the cooperation of multiple events. Transitions from one behaviour to another are possible over the range of stopping powers, and in some cases it is even possible to observe different regimes for similar stopping powers but different particle velocities\cite{fama2007}. The occurence of such complex transitions and other factors like absolute yields depend on the details of energy deposition and relaxation in the considered system.

Water ice has been explored on the broadest range of stopping power, and consistently exhibits a desorption yield proportional to the square of the stopping power\cite{dartois2015a}. This is also the case for CO ice\cite{chrisey1990,seperueloduarte2010}, and for CO$_2$ ice\cite{brown1982,seperueloduarte2009,raut2013,mejia2015}. The stopping power dependence of Ar ice is almost linear\cite{reimann1988}. For N$_2$ ice, a transition is observed from a quadratic dependence at high stopping power to a linear dependence at low stopping power\cite{johnson1991}. This particular transition has been related to a progressive increase of deposited energy density leading to an overlap of initially separated cascades triggered by single events.

\begin{figure}
	\centering
    \includegraphics[trim={0cm 0cm 0cm 0cm},clip,width=\linewidth]{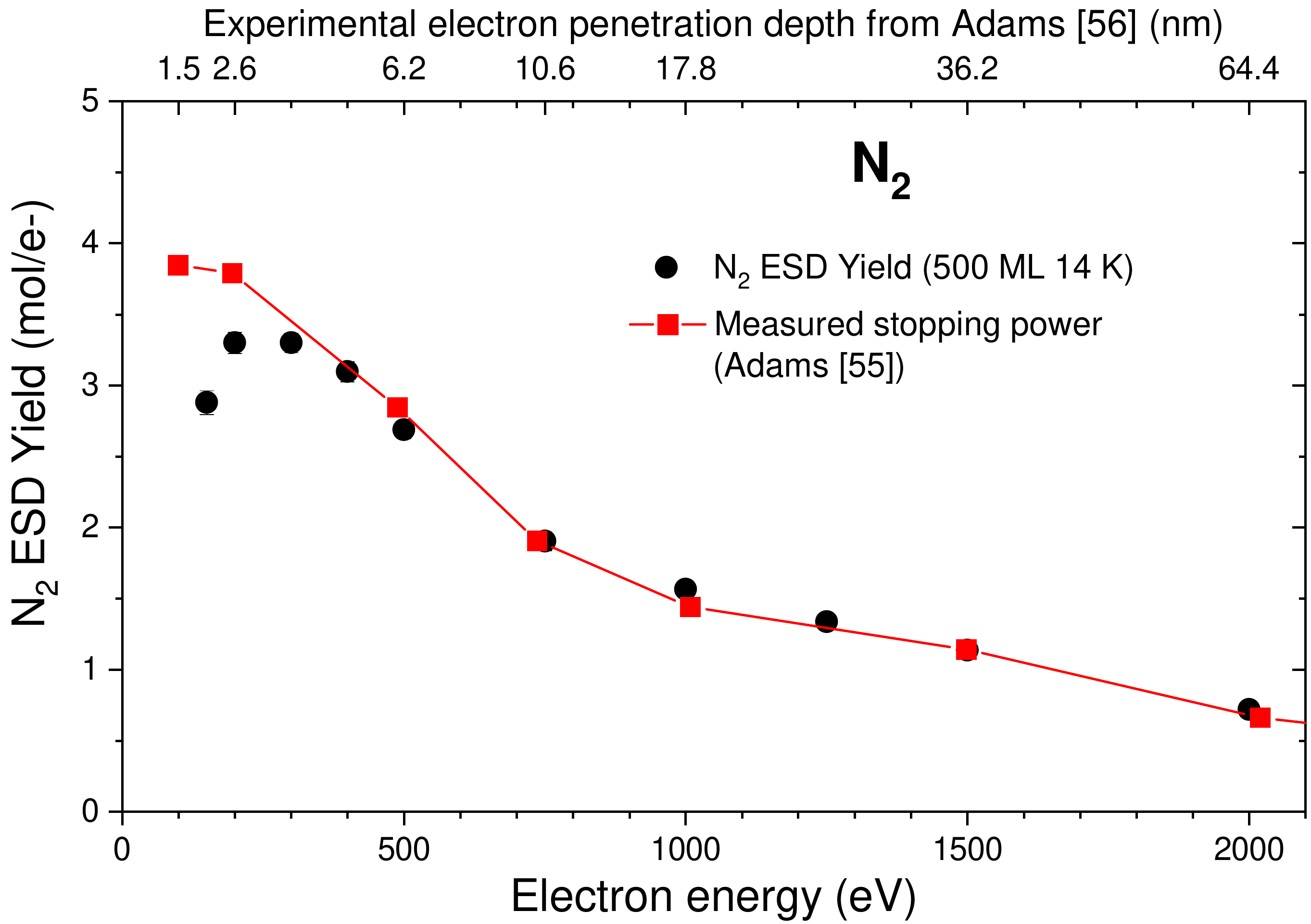}
    \caption{ESD yield of N$_2$ from 500 ML N$_2$ ice grown and irradiated at 14 K. Also shown for comparison is the measured stopping power of N$_2$ ice from ref. \cite{adams1980}, scaled to fit the high energy ESD curve.} 
    \label{N2_EID}
\end{figure}

\begin{figure}
	\centering
    \includegraphics[trim={0cm 0cm 0cm 0cm},clip,width=\linewidth]{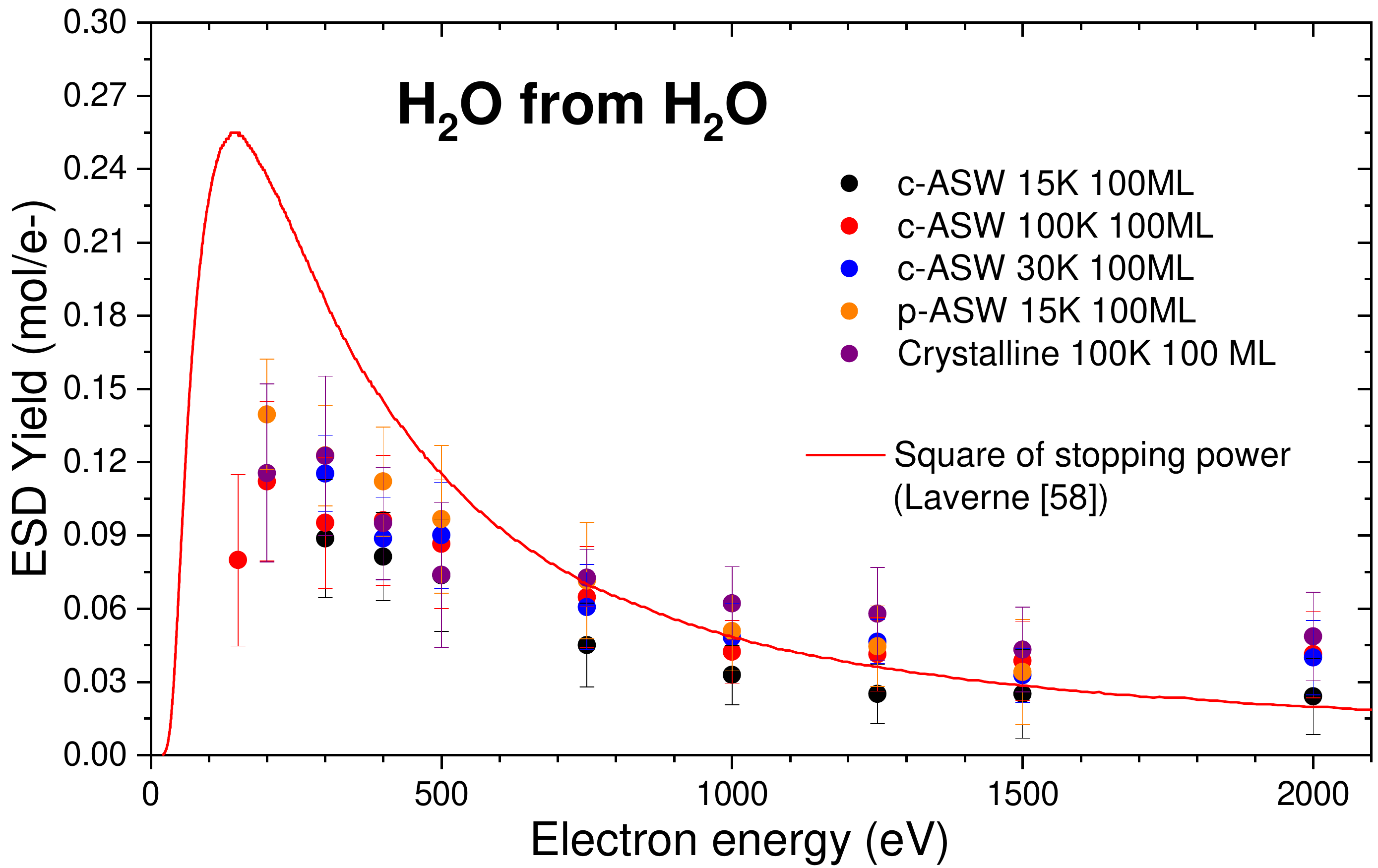}
    \caption{ESD yield of H$_2$O from different H$_2$O ices, similar to fig. \ref{H2O_D2O}a, compared with the square of the stopping power for water ice. The stopping power is adapted from ref. \cite{laverne1983}.} 
    \label{H2O_stopping}
\end{figure}

These stopping power dependences, which have most often been measured for ion sputtering, remain true in the keV electron regime (which corresponds to a low stopping power regime relative to most of the studies made with ions). In the sub-keV regime, on the other hand, we start encountering discrepancies between stopping power and ESD yields. Existing measurements of stopping power at low electron energies\cite{adams1980,gumus2005} show that stopping power keeps increasing when going towards lower electron energies, down to around 100 eV - similarly to electron-impact ionization cross-sections, for example, which are also related to the degree of electron-matter interaction. We saw in the various examples presented here that ESD curves on the other hand tend to have a maximum around 200-300 eV or even higher. For example, in fig. \ref{N2_EID}, the ESD yield of N$_2$ for a 500 ML ice grown and irradiated at 14 K, previously shown in fig. \ref{N2_EID_thickness} is compared to the measured stopping power of ref. \cite{adams1980} for electrons going through N$_2$ ice. We see a good agreement at high energies, in particular with the measured stopping power, but deviations at low energies. Another example is shown in fig. \ref{H2O_stopping} where the ESD curves of H$_2$O desorption from H$_2$O ices are compared with the square of the stopping power for the ice. The deviation at low energies is again clear. 

\begin{figure*}
	\centering
    \includegraphics[trim={0cm 3.3cm 0cm 0cm},clip,width=\linewidth]{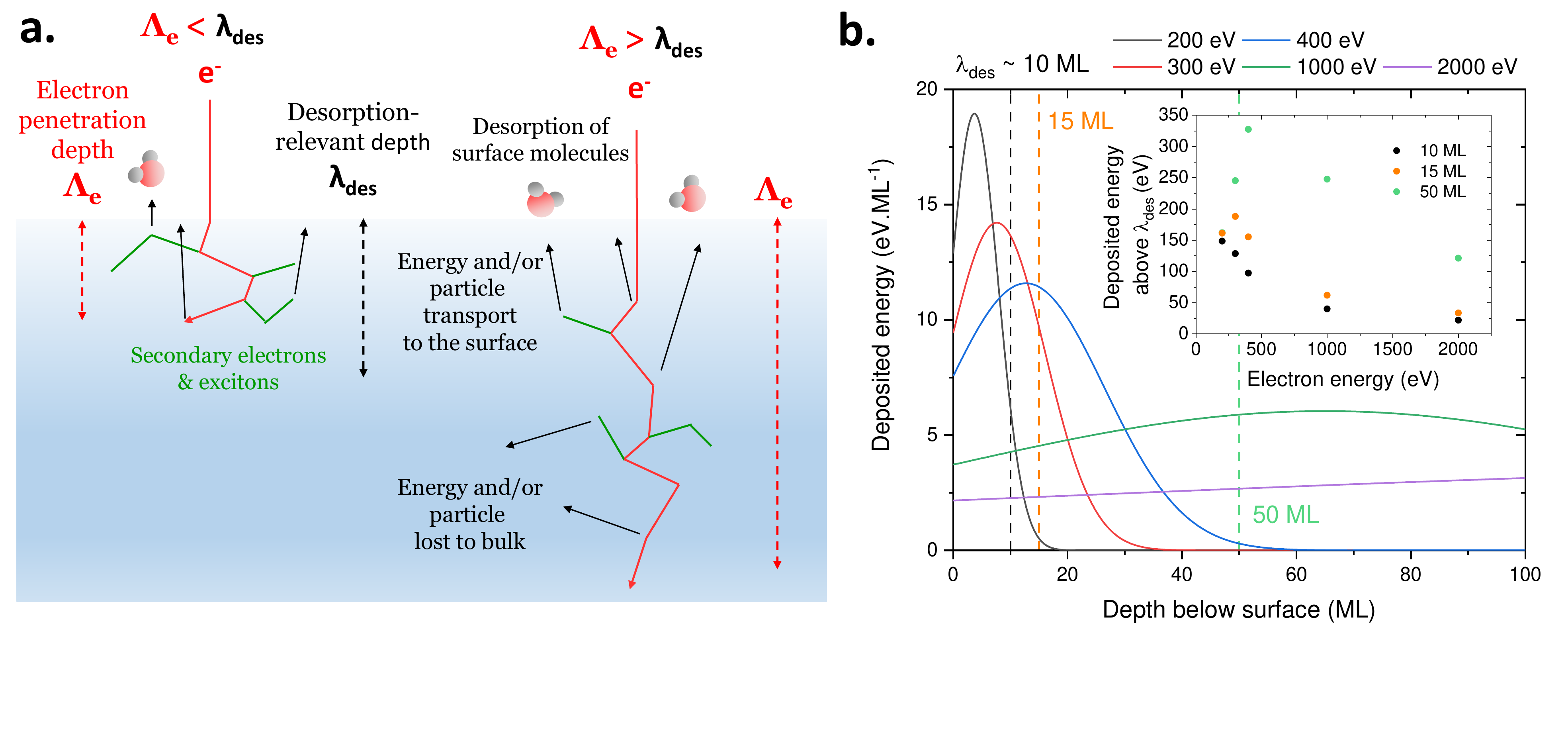}
    \caption{\textbf{a.} Schematic representation of the two energy deposition regimes for desorption induced by sub-keV electrons, delineated by the desorption-relevant depth (left-side: low energy electrons, right side: high energy electrons). \textbf{b.} Simulated energy deposition profiles of electrons of different energies in N$_2$ ice, using the parameters and formulas described in ref. \onlinecite{valkealahti1989}. Inset: integral of the deposited energy from 0 to \textlambda$_{des}$ = 10, 15 or 50 ML, as a function of electron energy.} 
    \label{DRD}
\end{figure*}

Deviations at low energy originate from the fact that the desorption yield is no longer proportional to the stopping power, because energy deposition occurs entirely near the surface. When the electron penetration depth becomes low enough, one needs first to take into account the details of the energy deposition structure, which is not exactly uniform. According to the Monte-Carlo simulations of ref.\onlinecite{valkealahti1989}, the energy deposition profile as a function of depth is better approximated by a truncated Gaussian. Examples at different electron energies were computed using the parameters given in ref. \onlinecite{valkealahti1989} for N$_2$ ice and are plotted in fig. \ref{DRD}b. But in addition to the exact energy deposition structure, one also needs to take into account energy (or particle) transport phenomena.

While desorption is really a surface process in the sense that the desorbing molecules are from the top molecular layer, a deeper part of the uppermost layers can be involved in desorption (see the example of the detailed study of ref. \onlinecite{bertin2012} for photon-stimulated desorption). Energy (for example in the form of excitons or electrons) or particles (energetic fragments, diffusing light species) can be transported from below up to the surface and may cause desorption, depending on the desorption mechanism involved. Such transport can be characterized by a length scale, which we call the desorption-relevant depth $\lambda_{des}$. This desorption-relevant depth corresponds to the transport length scale of the mechanism(s) which dominate energy and/or particle transport, and its exact definition will depend on these mechanisms.

For incident fast charged particles that have a penetration depth ($\Lambda_e$) in the ice much greater than the desorption-relevant depth, the desorption yield is related (linearly or by a higher order relation depending on the dominant type of desorption mechanism) to the stopping power, as particles all deposit uniformly within the whole desorption-relevant depth region an amount of energy proportional to that stopping power. However, when the penetration depth becomes comparable to the desorption-relevant depth, this will no longer be the case. The other extreme would be the case of incident electrons depositing all of their energy very close to the surface. We would then expect the desorption yield to be proportional to the energy of the electron, rather than its energy loss per unit distance. These two different regimes are schematized in fig. \ref{DRD}a. 

The amount of deposited energy within the desorption-relevant depth region will reach a maximum for the intermediate of these two regimes. In fig. \ref{DRD}b, the computed energy deposition profiles were integrated up to given values of $\lambda_{des}$, and the result as a function of electron energy is plotted in the inset. One can see how the maximum occurs at different electron energies depending on $\lambda_{des}$. These curves are simplified models that do not reproduce actual ESD curves, as one also needs to take into account how the energy (or particle) transport is attenuated. For example, if we had a transport mechanism with a diffusion-like behaviour, the amount of energy that could migrate from a given layer to the surface would exponentially decrease with the distance to the surface. The desorption-relevant depth would then be the characteristic length scale of this exponential.  

\subsection{Application to argon ice}

\begin{figure}
	\centering
    \includegraphics[trim={0cm 0cm 0cm 0cm},clip,width=\linewidth]{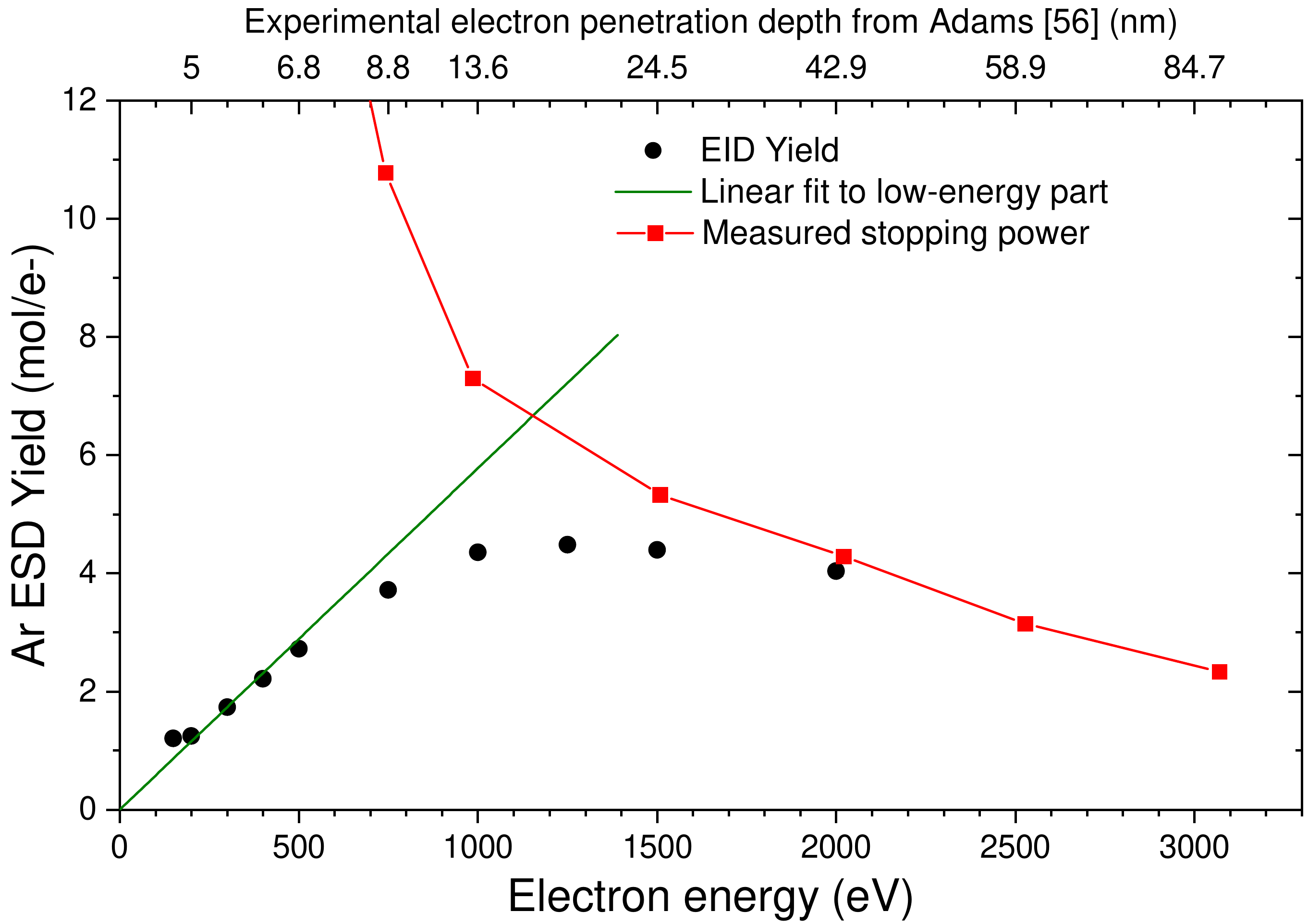}
    \caption{ESD yield of Ar from 500 ML Ar ice grown and irradiated at 14 K. The low energy part is fit to a linear relation (green line). Also shown for comparison is the measured stopping power of Ar ice from ref. \onlinecite{adams1980}, scaled to fit the high energy ESD curve.} 
    \label{Ar_EID}
\end{figure}

Argon ice provides useful demonstrations of the application of the notion of desorption-relevant depth. First, the above-mentioned low-energy linear regime is visible in the case of argon ice, where the transition between the two different regimes of energy deposition occurs at much higher electron energy than for the other ices. The ESD yield of a 500 ML Ar ice grown and irradiated at 14 K is presented on fig. \ref{Ar_EID}, along with the stopping power from ref. \onlinecite{adams1980}. Up until 500 eV the desorption yield is linear with electron energy (the lowest point at 150 eV is off because of charging issues). The regime in which the yield is proportional to stopping power is not visible in our electron energy range limited to 2 keV.

A consequence of the existence of these two regimes is that the ESD yield curve in the transition regime is characteristic of the desorption-relevant depth of the considered system. Around the maximum of the curve, where the ESD yield starts to deviate from the stopping power, we have $\lambda_{des} \sim \Lambda_e$ and it is therefore possible to estimate the value of $\lambda_{des}$, knowing the electron penetration depths. Hence for N$_2$ ice with a maximum around 250 eV we have $\lambda_{des} \sim$ 10 ML (30 \AA). In fig. \ref{N2_EID} and fig. \ref{Ar_EID} the upper axis shows the experimental estimates of the electron penetration depth as a function of electron energy from Adams et al. \cite{adams1980}. 

This principle is also well illustrated in the thickness dependence of Ar ice ESD, already presented in fig. \ref{Ar_EID_thickness}. So far an underlying assumption in our model was that the ices were considered semi-infinite. For ices of finite thickness, it is possible to have an ice thinner than the "intrinsic" desorption-relevant depth, in which case the whole ice is involved in desorption and the effective desorption-relevant depth becomes the ice thickness itself. This is what we observed for argon films in fig. \ref{Ar_EID_thickness}. The shape of the ESD yield curve changes with thickness: the maximum shifts progressively towards higher energies as the thickness increases, until the thickness of the ice becomes much higher than the intrinsic desorption-relevant depth, around 300 ML. From the ESD yield curve of the semi-infinite thickness ice, we can estimate that the desorption-relevant depth in our Argon ices is approximately 100 ML. This is in line with the saturation of the yields reached around 300 ML: for an exponentially attenuated energy migration process we would expect the yields to completely saturate above approximately 3 times the characteristic migration length scale.

This large intrinsic desorption-relevant depth is linked with the mechanisms of excitation transport in the ice, as mentioned previously. Desorption mechanisms and excitation transport in the case of rare-gas solids like argon have been extensively studied and are therefore well known. Rare gas solids host excitons that can migrate over large distances in the ice, with typical estimates in the 70 - 500 ML range depending on the studies \cite{zimmerer1994, ellegaard1988, reimann1988}, which is consistent with what we observe. Previous studies on the thickness dependence of electron or ion-stimulated desorption of argon found results very similar to ours \cite{reimann1988,ellegaard1988}. Migration of electronic excitations (excitons) is therefore what drives the desorption-relevant depth in the case of argon ice.

\begin{figure}
	\centering
    \includegraphics[trim={0cm 0cm 0cm 0cm},clip,width=\linewidth]{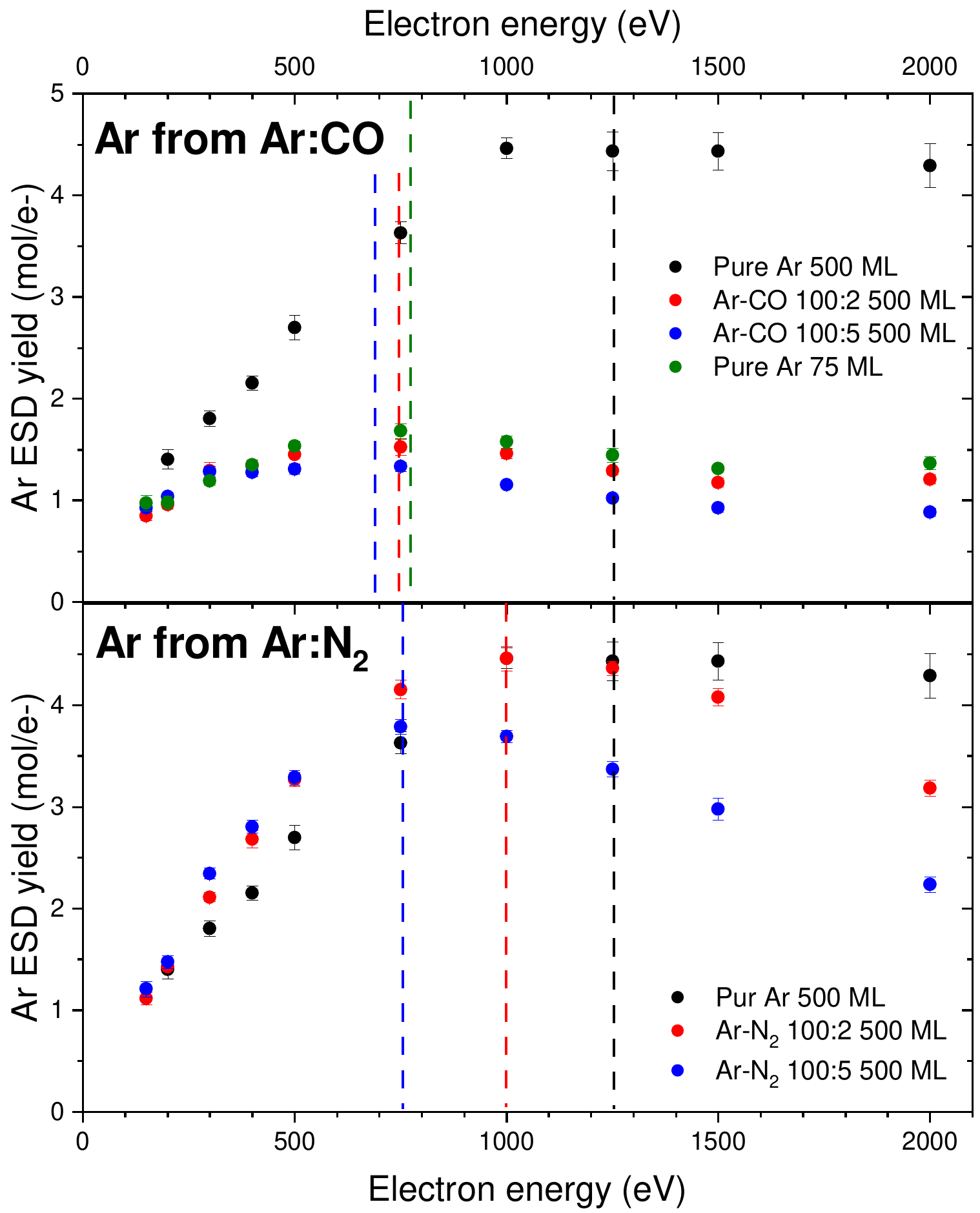}
    \caption{ESD yield of Ar from Ar ices with and without added impurities. In the upper panel the ESD yields from 500 ML Ar:CO ices with 2\% and 5\% CO are compared with ESD yields from 500 ML and 75 ML pure Ar ices. In the lower panel the ESD yields from 500 ML Ar:N$_2$ ices with 2\% and 5\% N$_2$ are compared with the ESD yield from 500 ML pure Ar ice.} 
    \label{Ar_imp_Ar}
\end{figure} 

The variation of the observed exciton migration distances in argon ice stems from its sensitivity to the presence of impurities and defects in the ice. To illustrate the influence of this on the desorption-relevant depth, and consequently on the ESD yields, we performed experiments where impurities of CO or N$_2$ were deliberately added to argon ice. The results are presented in fig. \ref{Ar_imp_Ar}. The Ar desorption yield for 500 ML ices of different compositions are presented. The upper panel corresponds to Ar:CO mixes with 2\% and 5\% CO, while the lower panel corresponds to Ar:N$_2$ mixes with similar amounts of N$_2$. The results are compared with the pure Ar case.

The addition of impurities significantly affects the ESD yield curves. In the case of CO, both curves are very different from the pure Ar 500 ML curve and closely resemble the 75 ML curve instead. This indicates that the CO impurities, acting as exciton sinks, effectively reduce the exciton migration distance, and therefore the desorption-relevant depth. The case of N$_2$ impurities (lower panel) is slightly different: while the curves display a different maximum, respectively around 1250, 1000 and 750 eV for the pure, 2\% and 5\% ices, the yields are higher than for the Ar:CO ice case or the pure 75 ML Ar ice case. We interpret this in the following way: the shape of the curves, in particular the position of the maximum, characterizes the desorption-relevant depth, which is again reduced by the addition of impurities. However, N$_2$ addition may also have an additional effect on the absolute Ar desorption yields, for example surface N$_2$ molecules may efficiently trap migrating excitons, and then transfer energy to neighboring Ar atoms, leading to desorption. The difference between N$_2$ and CO may reside in the electronic structure of the two molecules: the first dipole-allowed electronic state of solid CO is lower in energy (around 8 eV \cite{lu2005}) than the first solid Ar exciton (around 11.5 eV \cite{fugol1978}) while in the case of solid N$_2$ the first dipole-allowed state lies around 12 eV \cite{fayolle2013}, above the first exciton threshold. Thus energy transfer from CO to Ar should be hindered, preventing a boosting effect similar to the one observed for N$_2$. Complementary experiments using other types of impurities would be necessary to confirm these hypotheses. 

\subsection{Structure- and desorbate-dependent desorption-relevant depths in CO$_2$}

\begin{figure}
	\centering
    \includegraphics[trim={0cm 0cm 0cm 0cm},clip,width=\linewidth]{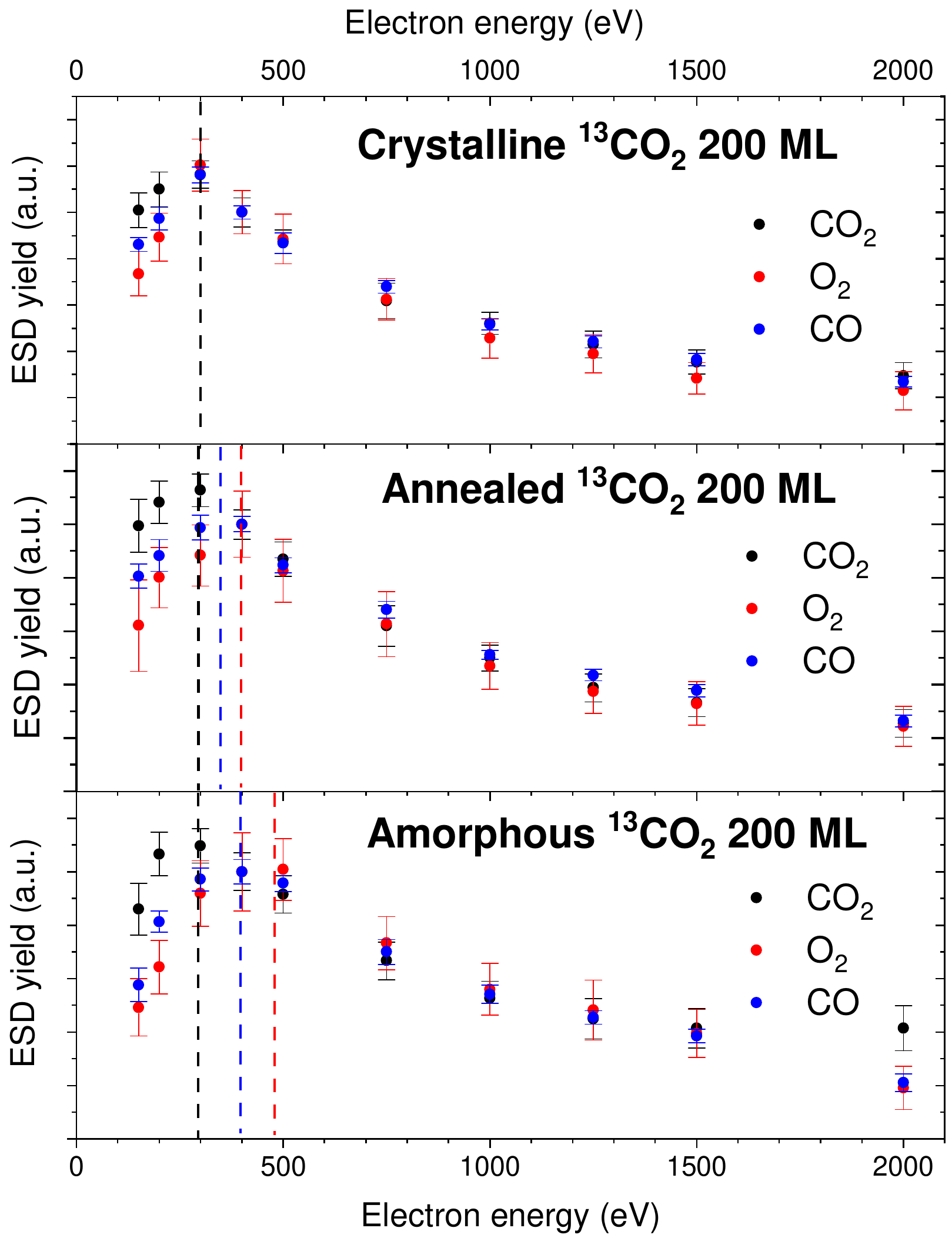}
    \caption{Comparison between the yields of CO, O$_2$ and CO$_2$ for the crystalline (top panel), the annealed (middle panel) and the porous amorphous (bottom panel) CO$_2$ ices. The yields have been normalized at 400 eV for comparison. The approximate maximum of each curve is indicated by dashed vertical lines.} 
    \label{CO2_phase}
\end{figure}

Let us now turn to the case of CO$_2$ ice. We recall that an interesting effect is observed in the desorption of CO and O$_2$, with a variation of their desorption maxima depending on the ice structure (fig. \ref{CO2_CO_O2_EID}). This effect can also be seen from a different angle by plotting instead the curves of the three desorbing products (normalized at 400 eV for comparison) for either the crystalline (grown at 75 K) ice, the annealed (grown at 14 K and annealed to 75 K) or the porous amorphous (grown at 14 K) ice, as in fig. \ref{CO2_phase}. In this figure we see that for the crystalline ice (upper panel), the maximum of the curve occurs basically at the same energy (300 eV) for all three species, despite some differences between the relative intensities. On the other hand, for the porous amorphous 14 K ice (lower panel), we see that the maximum is similar to the crystalline ice for CO$_2$ but shifts towards higher energies for CO and even more for O$_2$. The annealed ice (middle panel) is intermediary between amorphous and crystalline.

Considering the interpretation of the ESD yield curves developed above, where the maximum characterizes the desorption-relevant depth, this means that for the porous amorphous ice, deeper layers are involved in the desorption of O$_2$ than for CO, and for CO than for CO$_2$ - but this is not the case for the crystalline ice. This is well explained by considering the desorption mechanisms of these different species, especially the weight of chemistry (in the sense of exothermic (re)formation of these species from CO$_2$ dissociation products, and desorption due to the excess energy). Desorption of CO$_2$ can involve chemistry but also more direct processes. Desorption of CO is similar: it can desorb directly by dissociation of CO$_2$ but also form and desorb from chemical processes. As for O$_2$, the direct dissociation of CO$_2$ is not likely to yield it and therefore chemistry has to be involved in its formation and subsequent desorption. When the ice is porous and amorphous, radicals formed by dissociation of bulk CO$_2$ can reach the surface, react and enhance chemistry, which will affect the most, in order, O$_2$, then CO, then CO$_2$, considering the weight that chemistry should play in the formation and desorption of these three species. Note that diffusion is not necessarily thermal - the thermal diffusion of radicals in CO$_2$ ice is unknown but may be low at 14 K - but also non-thermal, due to the kinetic energy imparted to fragments upon dissociation. It is also possible that O$_2$ and CO formed in the bulk directly migrate to the surface, where they are subsequently desorbed by another excitation. These migration phenomena do not occur for crystalline CO$_2$ where bulk species cannot reach the surface (at least not at 14 K).

The desorption of e.g. O$_2$ from CO$_2$ ice is therefore an example of a case where the desorption-relevant depth is dominated not by energy transport (in the form of secondary electron or electronic excitation migration, for example) but particle transport: migration of species from the bulk to the surface. It also provides a good illustration of the use that ESD yield curves can have as a probe, revealing the efficiency of migration to the surface for porous amorphous versus crystalline ice. 

\begin{table*}
\centering
\def\arraystretch{1.2}
\caption{\label{summary} Summary of the results obtained on different molecular ices}
\begin{tabular}{>{\raggedright\arraybackslash}p{1.5cm} 
>{\centering\arraybackslash}p{2cm} 
>{\centering\arraybackslash}p{1.5cm} 
>{\centering\arraybackslash}p{2cm}
 >{\centering\arraybackslash}p{1cm}
>{\centering\arraybackslash}p{2cm} 
>{\centering\arraybackslash}p{3cm}}

Ice & Desorbing species & E$_{max}$ (eV) & Yield at E$_{max}$ (mol/electron) & n$^a$ & \textlambda$_{des}^b$ (ML) & Dominant source of transport \\
\hline
N$_2$ & N$_2$  & 250 & 3.4  & 1 & 10-15 & Secondary electrons \\
\hline
\multirow{2}{1.5cm}{CO} & CO     & 300     & 10.5 & 2 & 10-15 & \multirow{2}{3cm}{Secondary electrons / 
Diffusion} \\
   						  & CO$_2$ & 300-500 & 0.07 &   &  &  \\
\hline
Ar 						  & Ar & 1250 & 4.5  & 1 & 100 & Excitons \\
\hline
\multirow{3}{1.5cm}{CO$_2$ Crys.} & CO$_2$ & 300 & 0.38 & 2 & 10 & \multirow{3}{3cm}{Secondary electrons / Diffusion} \\
   								  & CO     & 300 & 1.9 &  & 10 &  \\
   								  & O$_2$  & 300 & 0.12 &  & 10 &  \\
\multirow{3}{1.5cm}{CO$_2$ Amorph.} & CO$_2$ & 300 & 0.38 & 2 & 10 & Secondary electrons / Diffusion \\
   									& CO     & 400 & 1.9  &  & 15 & Diffusion \\
   								   & O$_2$   & 500 & 0.12 &  & 20 & Diffusion  \\
\hline
\multirow{3}{1.5cm}{H$_2$O} & H$_2$O & 300-400 & 0.12 & 2 & 10-20 &  \\
   							   & H$_2$ & 300 &  &  &  &  \\
							   & O$_2$ & 300 &  &  &  &  \\
\multirow{3}{1.5cm}{D$_2$O} & D$_2$O & 200-300 & 0.022 & 2 & 10-20 &  \\
   							   & D$_2$ & 300 &  &  &  &  \\
							   & O$_2$ & 300 &  &  &  &  \\
\hline

\end{tabular} \\
$^a$ \footnotesize Exponent of the stopping power dependence in the keV electron regime \\
$^b$ \footnotesize Desorption-relevant depth estimated from the maxima of the ESD yield curves. All values are rough approximates (see the text for a detailed discussion).
\end{table*}

\subsection{Dominant contributions to the desorption-relevant depth of different molecular ices}

We finish by discussing the approximate desorption-relevant depth and the dominant type of transport setting this depth for the different molecular ices studied here. The desorption-relevant depth are estimated from the maximum of the ESD yield curves and the corresponding electron penetration depth. These only constitute rough estimations. The desorption-relevant depths are expressed here in MLs, and the approximate conversion from ML to \AA~is 1 ML $\sim$ 3 \AA~(the actual number depends on the ice density).

We addressed above the question of the desorption-relevant depth of argon ice, which is specific and different from the molecular ices as it hosts excitons. The desorption-relevant depth is $\lambda_{des} \sim$ 100 ML and is affected by the presence of impurities, as demonstrated in the experiments presented in fig. \ref{Ar_imp_Ar}. In the case of N$_2$ ice, we pointed out previously the similarity between the ESD curve and the SEY curve (fig. \ref{N2_EID_thickness}). The derived characteristic desorption-relevant depth is around 10-15 ML ($\sim$ 30 \AA), which is also the typical approximate escape depth of secondary electrons\cite{voreades1976,kuzucan2012}. Both of these facts point to the dominant energy carrier being secondary electrons in N$_2$ ice. The transport of secondary electrons is the default dominant contribution to the desorption-relevant depth in ESD, in the absence of other transport mechanisms with longer length scales, since it is always present. This also implies that transport mechanisms with length scales lower than $\sim$ 10 ML (such as the indirect desorption process identified in VUV PSD from CO ice\cite{bertin2012}) cannot be probed using this method.

For CO ice, there is a difference between SEY and ESD curves (fig. \ref{CO_EID}). Aside from the difference at high energy due to specular reflection of electrons, this difference can be explained by the square of stopping power dependence of ESD for this system. It is therefore not clear whether secondary electron transport is the only dominant mechanism here, although the desorption-relevant depth is similarly to N$_2$ around 10-15 ML. Chemistry leading to reformation and exothermic desorption of CO could contribute, and the quadratic stopping power dependence lends credit to this hypothesis, as desorption from exothermic chemistry involves multiple consecutive excitation events. There is no particular evidence for electronic energy transfer in CO ice. For CO$_2$ ice ($\lambda_{des} \sim$ 10 - 20 ML), we have shown that the desorption-relevant depth depends on the considered desorbing species and the structure of the ice. We explained such a difference by the diffusion of species from the bulk to the surface. As the desorption-relevant depth found is close to the escape depth of secondary electrons, they certainly also contribute in a significant way to transport in CO$_2$ ice.

For H$_2$O ice, experimental difficulties (high error bars, transitory behaviour of product desorption) obscure the conclusions that can be drawn. The ESD yield curves seem to have similar shapes within the experimental errors. We may have expected observations similar to CO$_2$ ice with maxima shifts due to diffusion in the porous structure. Instead the desorption of these species may be dominated by surface reactions, as emphasized in PSD studies \cite{dupuy2019e}. For the different structures and desorbing species from H$_2$O ice, $\lambda_{des}$ should not exceed 10-20 ML either considering the curve shapes. Mechanisms contributing to the desorption-relevant depth in H$_2$O ice could include, aside from secondary electrons, the diffusion of species and the presence of excitons that seem to have a migration length scale of the order of 20 ML\cite{petrik2005,marchione2016a}.

\section{Conclusion}

We quantified the electron-stimulated desorption of neutral species for various molecular ice systems (N$_2$, CO, CO$_2$, H$_2$O/D$_2$O) and solid Ar, in the 150 - 2000 eV electron energy range. Our results are in good agreement with the pre-existing quantified measurements of ESD on some of these systems, and expand to the desorption of species other than the parent molecule and the exploration of ice temperature and structure effects. These results are summarized in table \ref{summary}, which lists for the studied systems the desorbing species, the electron energy where the desorption yield is maximum, the corresponding maximum yield, as well as the above-discussed values of the desorption-relevant depths and corresponding dominant sources of transport in the ice. This new data as well as the physical considerations developed in this article and summarized below should help a better modelling of ESD-related phenomena in the above-mentioned fields of e.g. vacuum dynamics in accelerators or astrochemistry.

We showed that interpretation of the electron energy dependence of ESD in the sub-keV regime requires going beyond simple stopping power considerations. We introduce the notion of desorption-relevant depth, a quantity characterizing how transport of energy or particle from the bulk to the surface make layers below the surface contribute to desorption. This is illustrated in the specific case of Argon, where excitons allow for a particularly long transport range observed in the ice thickness dependence of the yield curve. For molecular ices, it also allows to evidence a difference of species migration to the surface in porous amorphous vs. crystalline phases, illustrated in the CO$_2$ ice case. In addition to bringing further fundamental understanding of the electron energy dependence of ESD, this study therefore also highlights the possible use of ESD as a probe of various phenomena (e.g. migration of electronic excitations or radicals and molecules) in systems like molecular ices.

\begin{acknowledgments}
This work was done under the collaboration agreement KE3324/TE between the European Organization for Nuclear Research (CERN) and Sorbonne Universit\'e, with financial support from CERN. The authors would like to thank M. Bertin, R. Cimino and G. F\'eraud for fruitful discussions on the topic of this manuscript and the presented results.
\end{acknowledgments}

\section*{Data availability}

The data that support the findings of this study are available from the corresponding author upon reasonable request.

%


\begin{thebibliography}{66}%
\makeatletter
\providecommand \@ifxundefined [1]{%
 \@ifx{#1\undefined}
}%
\providecommand \@ifnum [1]{%
 \ifnum #1\expandafter \@firstoftwo
 \else \expandafter \@secondoftwo
 \fi
}%
\providecommand \@ifx [1]{%
 \ifx #1\expandafter \@firstoftwo
 \else \expandafter \@secondoftwo
 \fi
}%
\providecommand \natexlab [1]{#1}%
\providecommand \enquote  [1]{``#1''}%
\providecommand \bibnamefont  [1]{#1}%
\providecommand \bibfnamefont [1]{#1}%
\providecommand \citenamefont [1]{#1}%
\providecommand \href@noop [0]{\@secondoftwo}%
\providecommand \href [0]{\begingroup \@sanitize@url \@href}%
\providecommand \@href[1]{\@@startlink{#1}\@@href}%
\providecommand \@@href[1]{\endgroup#1\@@endlink}%
\providecommand \@sanitize@url [0]{\catcode `\\12\catcode `\$12\catcode
  `\&12\catcode `\#12\catcode `\^12\catcode `\_12\catcode `\%12\relax}%
\providecommand \@@startlink[1]{}%
\providecommand \@@endlink[0]{}%
\providecommand \url  [0]{\begingroup\@sanitize@url \@url }%
\providecommand \@url [1]{\endgroup\@href {#1}{\urlprefix }}%
\providecommand \urlprefix  [0]{URL }%
\providecommand \Eprint [0]{\href }%
\providecommand \doibase [0]{https://doi.org/}%
\providecommand \selectlanguage [0]{\@gobble}%
\providecommand \bibinfo  [0]{\@secondoftwo}%
\providecommand \bibfield  [0]{\@secondoftwo}%
\providecommand \translation [1]{[#1]}%
\providecommand \BibitemOpen [0]{}%
\providecommand \bibitemStop [0]{}%
\providecommand \bibitemNoStop [0]{.\EOS\space}%
\providecommand \EOS [0]{\spacefactor3000\relax}%
\providecommand \BibitemShut  [1]{\csname bibitem#1\endcsname}%
\let\auto@bib@innerbib\@empty
\bibitem [{\citenamefont {Redhead}(1997)}]{redhead1997}%
  \BibitemOpen
  \bibfield  {author} {\bibinfo {author} {\bibfnamefont {P.~A.}\ \bibnamefont
  {Redhead}},\ }\bibfield  {title} {\enquote {\bibinfo {title} {The first 50
  years of electron stimulated desorption (1918–1968)},}\ }\href
  {https://doi.org/10.1016/S0042-207X(97)00030-4} {\bibfield  {journal}
  {\bibinfo  {journal} {Vacuum}\ }\textbf {\bibinfo {volume} {48}},\ \bibinfo
  {pages} {585--596} (\bibinfo {year} {1997})}\BibitemShut {NoStop}%
\bibitem [{\citenamefont {Dempster}(1918)}]{dempster1918}%
  \BibitemOpen
  \bibfield  {author} {\bibinfo {author} {\bibfnamefont {A.~J.}\ \bibnamefont
  {Dempster}},\ }\bibfield  {title} {\enquote {\bibinfo {title} {A new {Method}
  of {Positive} {Ray} {Analysis}},}\ }\href
  {https://doi.org/10.1103/PhysRev.11.316} {\bibfield  {journal} {\bibinfo
  {journal} {Phys. Rev.}\ }\textbf {\bibinfo {volume} {11}},\ \bibinfo {pages}
  {316--325} (\bibinfo {year} {1918})}\BibitemShut {NoStop}%
\bibitem [{\citenamefont {Menzel}(1995)}]{menzel1995a}%
  \BibitemOpen
  \bibfield  {author} {\bibinfo {author} {\bibfnamefont {D.}~\bibnamefont
  {Menzel}},\ }\bibfield  {title} {\enquote {\bibinfo {title} {Thirty years of
  {MGR}: {How} it came about, and what came of it},}\ }\href
  {https://doi.org/10.1016/0168-583X(95)00060-7} {\bibfield  {journal}
  {\bibinfo  {journal} {Nuclear Instruments and Methods in Physics Research
  Section B: Beam Interactions with Materials and Atoms}\ }\textbf {\bibinfo
  {volume} {101}},\ \bibinfo {pages} {1--10} (\bibinfo {year}
  {1995})}\BibitemShut {NoStop}%
\bibitem [{\citenamefont {Avouris}\ and\ \citenamefont
  {Walkup}(1989)}]{avouris1989}%
  \BibitemOpen
  \bibfield  {author} {\bibinfo {author} {\bibfnamefont {P.}~\bibnamefont
  {Avouris}}\ and\ \bibinfo {author} {\bibfnamefont {R.~E.}\ \bibnamefont
  {Walkup}},\ }\bibfield  {title} {\enquote {\bibinfo {title} {Fundamental
  mechanisms of desorption and fragmentation induced by electronic transitions
  at surfaces},}\ }\href {https://doi.org/10.1146/annurev.pc.40.100189.001133}
  {\bibfield  {journal} {\bibinfo  {journal} {Annual Review of Physical
  Chemistry}\ }\textbf {\bibinfo {volume} {40}},\ \bibinfo {pages} {173--206}
  (\bibinfo {year} {1989})}\BibitemShut {NoStop}%
\bibitem [{\citenamefont {Madey}(1994)}]{madey1994}%
  \BibitemOpen
  \bibfield  {author} {\bibinfo {author} {\bibfnamefont {T.~E.}\ \bibnamefont
  {Madey}},\ }\bibfield  {title} {\enquote {\bibinfo {title} {History of
  desorption induced by electronic transitions},}\ }\href
  {https://doi.org/10.1016/0039-6028(94)90700-5} {\bibfield  {journal}
  {\bibinfo  {journal} {Surface science}\ }\textbf {\bibinfo {volume} {299}},\
  \bibinfo {pages} {824--836} (\bibinfo {year} {1994})}\BibitemShut {NoStop}%
\bibitem [{\citenamefont {Gröbner}(1999)}]{grobner1999}%
  \BibitemOpen
  \bibfield  {author} {\bibinfo {author} {\bibfnamefont {O.}~\bibnamefont
  {Gröbner}},\ }\bibfield  {title} {\enquote {\bibinfo {title} {Dynamic
  {Outgassing}},}\ }in\ \href {https://doi.org/10.5170/CERN-1999-005.127}
  {\emph {\bibinfo {booktitle} {{CAS} - {CERN} {Accelerator} {School} and
  {ALBA} {Synchrotron} {Light} {Facility} : {Course} on {Vacuum} in
  {Accelerators}}}}\ (\bibinfo {year} {1999})\BibitemShut {NoStop}%
\bibitem [{\citenamefont {Hilleret}(2007)}]{hilleret2007}%
  \BibitemOpen
  \bibfield  {author} {\bibinfo {author} {\bibfnamefont {N.}~\bibnamefont
  {Hilleret}},\ }\bibfield  {title} {\enquote {\bibinfo {title} {Non-thermal
  outgassing},}\ }in\ \href {https://doi.org/10.5170/CERN-2007-003.87} {\emph
  {\bibinfo {booktitle} {{CAS} - {CERN} {Accelerator} {School} and {ALBA}
  {Synchrotron} {Light} {Facility} : {Course} on {Vacuum} in {Accelerators}}}}\
  (\bibinfo {year} {2007})\BibitemShut {NoStop}%
\bibitem [{\citenamefont {Cimino}\ and\ \citenamefont
  {Demma}(2014)}]{cimino2014a}%
  \BibitemOpen
  \bibfield  {author} {\bibinfo {author} {\bibfnamefont {R.}~\bibnamefont
  {Cimino}}\ and\ \bibinfo {author} {\bibfnamefont {T.}~\bibnamefont {Demma}},\
  }\bibfield  {title} {\enquote {\bibinfo {title} {Electron cloud in
  accelerators},}\ }\href {https://doi.org/10.1142/S0217751X14300233}
  {\bibfield  {journal} {\bibinfo  {journal} {Int. J. Mod. Phys. A}\ }\textbf
  {\bibinfo {volume} {29}},\ \bibinfo {pages} {1430023} (\bibinfo {year}
  {2014})},\ \bibinfo {note} {tex.ids: cimino2014}\BibitemShut {NoStop}%
\bibitem [{\citenamefont {Baglin}\ \emph {et~al.}(2002)\citenamefont {Baglin},
  \citenamefont {Collins}, \citenamefont {Gröbner}, \citenamefont
  {Grünhagel},\ and\ \citenamefont {Jenninger}}]{baglin2002}%
  \BibitemOpen
  \bibfield  {author} {\bibinfo {author} {\bibfnamefont {V.}~\bibnamefont
  {Baglin}}, \bibinfo {author} {\bibfnamefont {I.}~\bibnamefont {Collins}},
  \bibinfo {author} {\bibfnamefont {O.}~\bibnamefont {Gröbner}}, \bibinfo
  {author} {\bibfnamefont {C.}~\bibnamefont {Grünhagel}},\ and\ \bibinfo
  {author} {\bibfnamefont {B.}~\bibnamefont {Jenninger}},\ }\bibfield  {title}
  {\enquote {\bibinfo {title} {Molecular desorption by synchrotron radiation
  and sticking coefficient at cryogenic temperatures for {H2}, {CH4}, {CO} and
  {CO2}},}\ }\href {https://doi.org/10.1016/S0042-207X(02)00226-9} {\bibfield
  {journal} {\bibinfo  {journal} {Vacuum}\ }\textbf {\bibinfo {volume} {67}},\
  \bibinfo {pages} {421--428} (\bibinfo {year} {2002})}\BibitemShut {NoStop}%
\bibitem [{\citenamefont {Tratnik}, \citenamefont {Hilleret},\ and\
  \citenamefont {Störi}(2007)}]{tratnik2007}%
  \BibitemOpen
  \bibfield  {author} {\bibinfo {author} {\bibfnamefont {H.}~\bibnamefont
  {Tratnik}}, \bibinfo {author} {\bibfnamefont {N.}~\bibnamefont {Hilleret}},\
  and\ \bibinfo {author} {\bibfnamefont {H.}~\bibnamefont {Störi}},\
  }\bibfield  {title} {\enquote {\bibinfo {title} {The desorption of condensed
  noble gases and gas mixtures from cryogenic surfaces},}\ }\href
  {https://doi.org/10.1016/j.vacuum.2005.11.064} {\bibfield  {journal}
  {\bibinfo  {journal} {Vacuum}\ }\textbf {\bibinfo {volume} {81}},\ \bibinfo
  {pages} {731--737} (\bibinfo {year} {2007})}\BibitemShut {NoStop}%
\bibitem [{\citenamefont {Féraud}\ \emph {et~al.}(2019)\citenamefont
  {Féraud}, \citenamefont {Bertin}, \citenamefont {Romanzin}, \citenamefont
  {Dupuy}, \citenamefont {Le~Petit}, \citenamefont {Roueff}, \citenamefont
  {Philippe}, \citenamefont {Michaut}, \citenamefont {Jeseck},\ and\
  \citenamefont {Fillion}}]{feraud2019a}%
  \BibitemOpen
  \bibfield  {author} {\bibinfo {author} {\bibfnamefont {G.}~\bibnamefont
  {Féraud}}, \bibinfo {author} {\bibfnamefont {M.}~\bibnamefont {Bertin}},
  \bibinfo {author} {\bibfnamefont {C.}~\bibnamefont {Romanzin}}, \bibinfo
  {author} {\bibfnamefont {R.}~\bibnamefont {Dupuy}}, \bibinfo {author}
  {\bibfnamefont {F.}~\bibnamefont {Le~Petit}}, \bibinfo {author}
  {\bibfnamefont {E.}~\bibnamefont {Roueff}}, \bibinfo {author} {\bibfnamefont
  {L.}~\bibnamefont {Philippe}}, \bibinfo {author} {\bibfnamefont
  {X.}~\bibnamefont {Michaut}}, \bibinfo {author} {\bibfnamefont
  {P.}~\bibnamefont {Jeseck}},\ and\ \bibinfo {author} {\bibfnamefont {J.-H.}\
  \bibnamefont {Fillion}},\ }\bibfield  {title} {\enquote {\bibinfo {title}
  {Vacuum {Ultraviolet} {Photodesorption} and {Photofragmentation} of
  {Formaldehyde}-{Containing} {Ices}},}\ }\href
  {https://doi.org/10.1021/acsearthspacechem.9b00057} {\bibfield  {journal}
  {\bibinfo  {journal} {ACS Earth Space Chem.}\ }\textbf {\bibinfo {volume}
  {3}},\ \bibinfo {pages} {1135--1150} (\bibinfo {year} {2019})}\BibitemShut
  {NoStop}%
\bibitem [{\citenamefont {Dartois}\ \emph {et~al.}(2018)\citenamefont
  {Dartois}, \citenamefont {Chabot}, \citenamefont {Barkach}, \citenamefont
  {Rothard}, \citenamefont {Augé}, \citenamefont {Agnihotri}, \citenamefont
  {Domaracka},\ and\ \citenamefont {Boduch}}]{dartois2018}%
  \BibitemOpen
  \bibfield  {author} {\bibinfo {author} {\bibfnamefont {E.}~\bibnamefont
  {Dartois}}, \bibinfo {author} {\bibfnamefont {M.}~\bibnamefont {Chabot}},
  \bibinfo {author} {\bibfnamefont {T.~I.}\ \bibnamefont {Barkach}}, \bibinfo
  {author} {\bibfnamefont {H.}~\bibnamefont {Rothard}}, \bibinfo {author}
  {\bibfnamefont {B.}~\bibnamefont {Augé}}, \bibinfo {author} {\bibfnamefont
  {A.~N.}\ \bibnamefont {Agnihotri}}, \bibinfo {author} {\bibfnamefont
  {A.}~\bibnamefont {Domaracka}},\ and\ \bibinfo {author} {\bibfnamefont
  {P.}~\bibnamefont {Boduch}},\ }\bibfield  {title} {\enquote {\bibinfo {title}
  {Cosmic ray sputtering yield of interstellar {H2O} ice mantles : {Ice} mantle
  thickness dependence},}\ }\href {https://doi.org/10.1051/0004-6361/201833277}
  {\bibfield  {journal} {\bibinfo  {journal} {Astronomy \& Astrophysics}\
  }\textbf {\bibinfo {volume} {618}},\ \bibinfo {pages} {A173} (\bibinfo {year}
  {2018})},\ \bibinfo {note} {arXiv: 1809.09180}\BibitemShut {NoStop}%
\bibitem [{\citenamefont {Brown}\ \emph {et~al.}(1984)\citenamefont {Brown},
  \citenamefont {Augustyniak}, \citenamefont {Marcantonio}, \citenamefont
  {Simmons}, \citenamefont {Boring}, \citenamefont {Johnson},\ and\
  \citenamefont {Reimann}}]{brown1984}%
  \BibitemOpen
  \bibfield  {author} {\bibinfo {author} {\bibfnamefont {W.~L.}\ \bibnamefont
  {Brown}}, \bibinfo {author} {\bibfnamefont {W.~M.}\ \bibnamefont
  {Augustyniak}}, \bibinfo {author} {\bibfnamefont {K.~J.}\ \bibnamefont
  {Marcantonio}}, \bibinfo {author} {\bibfnamefont {E.~H.}\ \bibnamefont
  {Simmons}}, \bibinfo {author} {\bibfnamefont {J.~W.}\ \bibnamefont {Boring}},
  \bibinfo {author} {\bibfnamefont {R.~E.}\ \bibnamefont {Johnson}},\ and\
  \bibinfo {author} {\bibfnamefont {C.~T.}\ \bibnamefont {Reimann}},\
  }\bibfield  {title} {\enquote {\bibinfo {title} {Electronic sputtering of low
  temperature molecular solids},}\ }\href
  {https://doi.org/10.1016/0168-583X(84)90085-5} {\bibfield  {journal}
  {\bibinfo  {journal} {Nuclear Instruments and Methods in Physics Research
  Section B: Beam Interactions with Materials and Atoms}\ }\textbf {\bibinfo
  {volume} {1}},\ \bibinfo {pages} {307--314} (\bibinfo {year}
  {1984})}\BibitemShut {NoStop}%
\bibitem [{\citenamefont {Muñoz~Caro}\ and\ \citenamefont
  {Martín~Doménech}(2018)}]{munozcaro2018a}%
  \BibitemOpen
  \bibfield  {author} {\bibinfo {author} {\bibfnamefont {G.~M.}\ \bibnamefont
  {Muñoz~Caro}}\ and\ \bibinfo {author} {\bibfnamefont {R.}~\bibnamefont
  {Martín~Doménech}},\ }\bibfield  {title} {\enquote {\bibinfo {title}
  {Photon-{Induced} {Desorption} {Processes} in {Astrophysical} {Ices}},}\ }in\
  \href {https://doi.org/10.1007/978-3-319-90020-9_9} {\emph {\bibinfo
  {booktitle} {Laboratory {Astrophysics}}}},\ Vol.\ \bibinfo {volume} {451},\
  \bibinfo {editor} {edited by\ \bibinfo {editor} {\bibfnamefont {G.~M.}\
  \bibnamefont {Muñoz~Caro}}\ and\ \bibinfo {editor} {\bibfnamefont
  {R.}~\bibnamefont {Escribano}}}\ (\bibinfo  {publisher} {Springer
  International Publishing},\ \bibinfo {address} {Cham},\ \bibinfo {year}
  {2018})\ pp.\ \bibinfo {pages} {133--147}\BibitemShut {NoStop}%
\bibitem [{\citenamefont {Dupuy}\ \emph {et~al.}(2018)\citenamefont {Dupuy},
  \citenamefont {Bertin}, \citenamefont {Féraud}, \citenamefont {Hassenfratz},
  \citenamefont {Michaut}, \citenamefont {Putaud}, \citenamefont {Philippe},
  \citenamefont {Jeseck}, \citenamefont {Angelucci}, \citenamefont {Cimino},
  \citenamefont {Baglin}, \citenamefont {Romanzin},\ and\ \citenamefont
  {Fillion}}]{dupuy2018c}%
  \BibitemOpen
  \bibfield  {author} {\bibinfo {author} {\bibfnamefont {R.}~\bibnamefont
  {Dupuy}}, \bibinfo {author} {\bibfnamefont {M.}~\bibnamefont {Bertin}},
  \bibinfo {author} {\bibfnamefont {G.}~\bibnamefont {Féraud}}, \bibinfo
  {author} {\bibfnamefont {M.}~\bibnamefont {Hassenfratz}}, \bibinfo {author}
  {\bibfnamefont {X.}~\bibnamefont {Michaut}}, \bibinfo {author} {\bibfnamefont
  {T.}~\bibnamefont {Putaud}}, \bibinfo {author} {\bibfnamefont
  {L.}~\bibnamefont {Philippe}}, \bibinfo {author} {\bibfnamefont
  {P.}~\bibnamefont {Jeseck}}, \bibinfo {author} {\bibfnamefont
  {M.}~\bibnamefont {Angelucci}}, \bibinfo {author} {\bibfnamefont
  {R.}~\bibnamefont {Cimino}}, \bibinfo {author} {\bibfnamefont
  {V.}~\bibnamefont {Baglin}}, \bibinfo {author} {\bibfnamefont
  {C.}~\bibnamefont {Romanzin}},\ and\ \bibinfo {author} {\bibfnamefont
  {J.-H.}\ \bibnamefont {Fillion}},\ }\bibfield  {title} {\enquote {\bibinfo
  {title} {X-ray photodesorption from water ice in protoplanetary disks and
  {X}-ray-dominated regions},}\ }\href
  {https://doi.org/10.1038/s41550-018-0532-y} {\bibfield  {journal} {\bibinfo
  {journal} {Nature Astronomy}\ }\textbf {\bibinfo {volume} {2}},\ \bibinfo
  {pages} {796--801} (\bibinfo {year} {2018})}\BibitemShut {NoStop}%
\bibitem [{\citenamefont {Johnson}\ \emph {et~al.}(2013)\citenamefont
  {Johnson}, \citenamefont {Carlson}, \citenamefont {Cassidy},\ and\
  \citenamefont {Fama}}]{johnson2013}%
  \BibitemOpen
  \bibfield  {author} {\bibinfo {author} {\bibfnamefont {R.~E.}\ \bibnamefont
  {Johnson}}, \bibinfo {author} {\bibfnamefont {R.~W.}\ \bibnamefont
  {Carlson}}, \bibinfo {author} {\bibfnamefont {T.~A.}\ \bibnamefont
  {Cassidy}},\ and\ \bibinfo {author} {\bibfnamefont {M.}~\bibnamefont
  {Fama}},\ }\bibfield  {title} {\enquote {\bibinfo {title} {Sputtering of
  {Ices}},}\ }in\ \href {http://link.springer.com/10.1007/978-1-4614-3076-6_17}
  {\emph {\bibinfo {booktitle} {The {Science} of {Solar} {System} {Ices}}}},\
  Vol.\ \bibinfo {volume} {356},\ \bibinfo {editor} {edited by\ \bibinfo
  {editor} {\bibfnamefont {M.~S.}\ \bibnamefont {Gudipati}}\ and\ \bibinfo
  {editor} {\bibfnamefont {J.}~\bibnamefont {Castillo-Rogez}}}\ (\bibinfo
  {publisher} {Springer New York},\ \bibinfo {address} {New York, NY},\
  \bibinfo {year} {2013})\ pp.\ \bibinfo {pages} {551--581}\BibitemShut
  {NoStop}%
\bibitem [{\citenamefont {Ellegaard}\ \emph {et~al.}(1986)\citenamefont
  {Ellegaard}, \citenamefont {Schou}, \citenamefont {Sørensen},\ and\
  \citenamefont {Børgesen}}]{ellegaard1986}%
  \BibitemOpen
  \bibfield  {author} {\bibinfo {author} {\bibfnamefont {O.}~\bibnamefont
  {Ellegaard}}, \bibinfo {author} {\bibfnamefont {J.}~\bibnamefont {Schou}},
  \bibinfo {author} {\bibfnamefont {H.}~\bibnamefont {Sørensen}},\ and\
  \bibinfo {author} {\bibfnamefont {P.}~\bibnamefont {Børgesen}},\ }\bibfield
  {title} {\enquote {\bibinfo {title} {Electronic sputtering of solid nitrogen
  and oxygen by {keV} electrons},}\ }\href
  {https://doi.org/10.1016/0039-6028(86)90718-1} {\bibfield  {journal}
  {\bibinfo  {journal} {Surface science}\ }\textbf {\bibinfo {volume} {167}},\
  \bibinfo {pages} {474--492} (\bibinfo {year} {1986})}\BibitemShut {NoStop}%
\bibitem [{\citenamefont {Rothard}\ \emph {et~al.}(2017)\citenamefont
  {Rothard}, \citenamefont {Domaracka}, \citenamefont {Boduch}, \citenamefont
  {Palumbo}, \citenamefont {Strazzulla}, \citenamefont {da~Silveira},\ and\
  \citenamefont {Dartois}}]{rothard2017}%
  \BibitemOpen
  \bibfield  {author} {\bibinfo {author} {\bibfnamefont {H.}~\bibnamefont
  {Rothard}}, \bibinfo {author} {\bibfnamefont {A.}~\bibnamefont {Domaracka}},
  \bibinfo {author} {\bibfnamefont {P.}~\bibnamefont {Boduch}}, \bibinfo
  {author} {\bibfnamefont {M.~E.}\ \bibnamefont {Palumbo}}, \bibinfo {author}
  {\bibfnamefont {G.}~\bibnamefont {Strazzulla}}, \bibinfo {author}
  {\bibfnamefont {E.~F.}\ \bibnamefont {da~Silveira}},\ and\ \bibinfo {author}
  {\bibfnamefont {E.}~\bibnamefont {Dartois}},\ }\bibfield  {title} {\enquote
  {\bibinfo {title} {Modification of ices by cosmic rays and solar wind},}\
  }\href {https://doi.org/10.1088/1361-6455/50/6/062001} {\bibfield  {journal}
  {\bibinfo  {journal} {Journal of Physics B: Atomic, Molecular and Optical
  Physics}\ }\textbf {\bibinfo {volume} {50}},\ \bibinfo {pages} {062001}
  (\bibinfo {year} {2017})},\ \bibinfo {note} {tex.ids:
  rothard2017a}\BibitemShut {NoStop}%
\bibitem [{\citenamefont {Boduch}\ \emph {et~al.}(2015)\citenamefont {Boduch},
  \citenamefont {Dartois}, \citenamefont {de~Barros}, \citenamefont
  {da~Silveira}, \citenamefont {Domaracka}, \citenamefont {Lv}, \citenamefont
  {Palumbo}, \citenamefont {Pilling}, \citenamefont {Rothard}, \citenamefont
  {Duarte},\ and\ \citenamefont {Strazzulla}}]{boduch2015}%
  \BibitemOpen
  \bibfield  {author} {\bibinfo {author} {\bibfnamefont {P.}~\bibnamefont
  {Boduch}}, \bibinfo {author} {\bibfnamefont {E.}~\bibnamefont {Dartois}},
  \bibinfo {author} {\bibfnamefont {A.~L.~F.}\ \bibnamefont {de~Barros}},
  \bibinfo {author} {\bibfnamefont {E.~F.}\ \bibnamefont {da~Silveira}},
  \bibinfo {author} {\bibfnamefont {A.}~\bibnamefont {Domaracka}}, \bibinfo
  {author} {\bibfnamefont {X.-Y.}\ \bibnamefont {Lv}}, \bibinfo {author}
  {\bibfnamefont {M.~E.}\ \bibnamefont {Palumbo}}, \bibinfo {author}
  {\bibfnamefont {S.}~\bibnamefont {Pilling}}, \bibinfo {author} {\bibfnamefont
  {H.}~\bibnamefont {Rothard}}, \bibinfo {author} {\bibfnamefont {E.~S.}\
  \bibnamefont {Duarte}},\ and\ \bibinfo {author} {\bibfnamefont
  {G.}~\bibnamefont {Strazzulla}},\ }\bibfield  {title} {\enquote {\bibinfo
  {title} {Radiation effects in astrophysical ices},}\ }\href
  {https://doi.org/10.1088/1742-6596/629/1/012008} {\bibfield  {journal}
  {\bibinfo  {journal} {J. Phys.: Conf. Ser.}\ }\textbf {\bibinfo {volume}
  {629}},\ \bibinfo {pages} {012008} (\bibinfo {year} {2015})}\BibitemShut
  {NoStop}%
\bibitem [{\citenamefont {Dartois}\ \emph {et~al.}(2015)\citenamefont
  {Dartois}, \citenamefont {Augé}, \citenamefont {Rothard}, \citenamefont
  {Boduch}, \citenamefont {Brunetto}, \citenamefont {Chabot}, \citenamefont
  {Domaracka}, \citenamefont {Ding}, \citenamefont {Kamalou}, \citenamefont
  {Lv}, \citenamefont {da~Silveira}, \citenamefont {Thomas}, \citenamefont
  {Pino}, \citenamefont {Mejia}, \citenamefont {Godard},\ and\ \citenamefont
  {de~Barros}}]{dartois2015a}%
  \BibitemOpen
  \bibfield  {author} {\bibinfo {author} {\bibfnamefont {E.}~\bibnamefont
  {Dartois}}, \bibinfo {author} {\bibfnamefont {B.}~\bibnamefont {Augé}},
  \bibinfo {author} {\bibfnamefont {H.}~\bibnamefont {Rothard}}, \bibinfo
  {author} {\bibfnamefont {P.}~\bibnamefont {Boduch}}, \bibinfo {author}
  {\bibfnamefont {R.}~\bibnamefont {Brunetto}}, \bibinfo {author}
  {\bibfnamefont {M.}~\bibnamefont {Chabot}}, \bibinfo {author} {\bibfnamefont
  {A.}~\bibnamefont {Domaracka}}, \bibinfo {author} {\bibfnamefont {J.-J.}\
  \bibnamefont {Ding}}, \bibinfo {author} {\bibfnamefont {O.}~\bibnamefont
  {Kamalou}}, \bibinfo {author} {\bibfnamefont {X.-Y.}\ \bibnamefont {Lv}},
  \bibinfo {author} {\bibfnamefont {E.~F.}\ \bibnamefont {da~Silveira}},
  \bibinfo {author} {\bibfnamefont {J.-C.}\ \bibnamefont {Thomas}}, \bibinfo
  {author} {\bibfnamefont {T.}~\bibnamefont {Pino}}, \bibinfo {author}
  {\bibfnamefont {C.}~\bibnamefont {Mejia}}, \bibinfo {author} {\bibfnamefont
  {M.}~\bibnamefont {Godard}},\ and\ \bibinfo {author} {\bibfnamefont
  {A.}~\bibnamefont {de~Barros}},\ }\bibfield  {title} {\enquote {\bibinfo
  {title} {Swift heavy ion modifications of astrophysical water ice},}\ }\href
  {https://doi.org/10.1016/j.nimb.2015.08.053} {\bibfield  {journal} {\bibinfo
  {journal} {Nuclear Instruments and Methods in Physics Research Section B:
  Beam Interactions with Materials and Atoms}\ }\textbf {\bibinfo {volume}
  {365}},\ \bibinfo {pages} {472--476} (\bibinfo {year} {2015})}\BibitemShut
  {NoStop}%
\bibitem [{\citenamefont {Huang}\ \emph {et~al.}(2020)\citenamefont {Huang},
  \citenamefont {Ciaravella}, \citenamefont {Cecchi-Pestellini}, \citenamefont
  {Jiménez-Escobar}, \citenamefont {Hsiao}, \citenamefont {Huang},
  \citenamefont {Chen}, \citenamefont {Sie},\ and\ \citenamefont
  {Chen}}]{huang2020}%
  \BibitemOpen
  \bibfield  {author} {\bibinfo {author} {\bibfnamefont {C.-H.}\ \bibnamefont
  {Huang}}, \bibinfo {author} {\bibfnamefont {A.}~\bibnamefont {Ciaravella}},
  \bibinfo {author} {\bibfnamefont {C.}~\bibnamefont {Cecchi-Pestellini}},
  \bibinfo {author} {\bibfnamefont {A.}~\bibnamefont {Jiménez-Escobar}},
  \bibinfo {author} {\bibfnamefont {L.-C.}\ \bibnamefont {Hsiao}}, \bibinfo
  {author} {\bibfnamefont {C.-C.}\ \bibnamefont {Huang}}, \bibinfo {author}
  {\bibfnamefont {P.-C.}\ \bibnamefont {Chen}}, \bibinfo {author}
  {\bibfnamefont {N.-E.}\ \bibnamefont {Sie}},\ and\ \bibinfo {author}
  {\bibfnamefont {Y.-J.}\ \bibnamefont {Chen}},\ }\bibfield  {title} {\enquote
  {\bibinfo {title} {Effects of 150–1000 {eV} {Electron} {Impacts} on {Pure}
  {Carbon} {Monoxide} {Ices} {Using} the {Interstellar} {Energetic}-{Process}
  {System} ({IEPS})},}\ }\href {https://doi.org/10.3847/1538-4357/ab5dbe}
  {\bibfield  {journal} {\bibinfo  {journal} {ApJ}\ }\textbf {\bibinfo {volume}
  {889}},\ \bibinfo {pages} {57} (\bibinfo {year} {2020})},\ \bibinfo {note}
  {tex.ids: huang2019 arXiv: 1912.11820}\BibitemShut {NoStop}%
\bibitem [{\citenamefont {Doronin}\ \emph {et~al.}(2015)\citenamefont
  {Doronin}, \citenamefont {Bertin}, \citenamefont {Michaut}, \citenamefont
  {Philippe},\ and\ \citenamefont {Fillion}}]{doronin2015}%
  \BibitemOpen
  \bibfield  {author} {\bibinfo {author} {\bibfnamefont {M.}~\bibnamefont
  {Doronin}}, \bibinfo {author} {\bibfnamefont {M.}~\bibnamefont {Bertin}},
  \bibinfo {author} {\bibfnamefont {X.}~\bibnamefont {Michaut}}, \bibinfo
  {author} {\bibfnamefont {L.}~\bibnamefont {Philippe}},\ and\ \bibinfo
  {author} {\bibfnamefont {J.-H.}\ \bibnamefont {Fillion}},\ }\bibfield
  {title} {\enquote {\bibinfo {title} {Adsorption energies and prefactor
  determination for {CH3OH} adsorption on graphite},}\ }\href
  {https://doi.org/10.1063/1.4929376} {\bibfield  {journal} {\bibinfo
  {journal} {The Journal of Chemical Physics}\ }\textbf {\bibinfo {volume}
  {143}},\ \bibinfo {pages} {084703} (\bibinfo {year} {2015})}\BibitemShut
  {NoStop}%
\bibitem [{\citenamefont {Feulner}\ and\ \citenamefont
  {Menzel}(1980)}]{feulner1980}%
  \BibitemOpen
  \bibfield  {author} {\bibinfo {author} {\bibfnamefont {P.}~\bibnamefont
  {Feulner}}\ and\ \bibinfo {author} {\bibfnamefont {D.}~\bibnamefont
  {Menzel}},\ }\bibfield  {title} {\enquote {\bibinfo {title} {Simple ways to
  improve ’’flash desorption’’ measurements from single crystal
  surfaces},}\ }\href {https://doi.org/10.1116/1.570537} {\bibfield  {journal}
  {\bibinfo  {journal} {Journal of Vacuum Science and Technology}\ }\textbf
  {\bibinfo {volume} {17}},\ \bibinfo {pages} {662--663} (\bibinfo {year}
  {1980})}\BibitemShut {NoStop}%
\bibitem [{\citenamefont {Kuzucan}\ \emph {et~al.}(2012)\citenamefont
  {Kuzucan}, \citenamefont {Neupert}, \citenamefont {Taborelli},\ and\
  \citenamefont {Störi}}]{kuzucan2012}%
  \BibitemOpen
  \bibfield  {author} {\bibinfo {author} {\bibfnamefont {A.}~\bibnamefont
  {Kuzucan}}, \bibinfo {author} {\bibfnamefont {H.}~\bibnamefont {Neupert}},
  \bibinfo {author} {\bibfnamefont {M.}~\bibnamefont {Taborelli}},\ and\
  \bibinfo {author} {\bibfnamefont {H.}~\bibnamefont {Störi}},\ }\bibfield
  {title} {\enquote {\bibinfo {title} {Secondary electron yield on cryogenic
  surfaces as a function of physisorbed gases},}\ }\href
  {https://doi.org/10.1116/1.4736552} {\bibfield  {journal} {\bibinfo
  {journal} {Journal of Vacuum Science \& Technology A: Vacuum, Surfaces, and
  Films}\ }\textbf {\bibinfo {volume} {30}},\ \bibinfo {pages} {051401}
  (\bibinfo {year} {2012})}\BibitemShut {NoStop}%
\bibitem [{\citenamefont {Pedrys}\ \emph {et~al.}(1989)\citenamefont {Pedrys},
  \citenamefont {Oostra}, \citenamefont {Haring}, \citenamefont {Devries},\
  and\ \citenamefont {Schou}}]{pedrys1989}%
  \BibitemOpen
  \bibfield  {author} {\bibinfo {author} {\bibfnamefont {R.}~\bibnamefont
  {Pedrys}}, \bibinfo {author} {\bibfnamefont {D.~J.}\ \bibnamefont {Oostra}},
  \bibinfo {author} {\bibfnamefont {A.}~\bibnamefont {Haring}}, \bibinfo
  {author} {\bibfnamefont {A.~E.}\ \bibnamefont {Devries}},\ and\ \bibinfo
  {author} {\bibfnamefont {J.}~\bibnamefont {Schou}},\ }\bibfield  {title}
  {\enquote {\bibinfo {title} {Energy distributions from electron-sputtered
  solid nitrogen},}\ }\href {https://doi.org/10.1080/10420158908220538}
  {\bibfield  {journal} {\bibinfo  {journal} {Radiation Effects and Defects in
  Solids}\ }\textbf {\bibinfo {volume} {109}},\ \bibinfo {pages} {239--244}
  (\bibinfo {year} {1989})}\BibitemShut {NoStop}%
\bibitem [{\citenamefont {Vegard}(1929)}]{vegard1929}%
  \BibitemOpen
  \bibfield  {author} {\bibinfo {author} {\bibfnamefont {L.}~\bibnamefont
  {Vegard}},\ }\bibfield  {title} {\enquote {\bibinfo {title} {Die {Struktur}
  derjenigen {Form} von festem {Stickstoff}, die unterhalb 35,50 {K} stabil
  ist},}\ }\href {https://doi.org/10.1007/BF01342209} {\bibfield  {journal}
  {\bibinfo  {journal} {Zeitschrift für Physik}\ }\textbf {\bibinfo {volume}
  {58}},\ \bibinfo {pages} {497--510} (\bibinfo {year} {1929})}\BibitemShut
  {NoStop}%
\bibitem [{\citenamefont {Satorre}\ \emph {et~al.}(2008)\citenamefont
  {Satorre}, \citenamefont {Domingo}, \citenamefont {Millán}, \citenamefont
  {Luna}, \citenamefont {Vilaplana},\ and\ \citenamefont
  {Santonja}}]{satorre2008}%
  \BibitemOpen
  \bibfield  {author} {\bibinfo {author} {\bibfnamefont {M.}~\bibnamefont
  {Satorre}}, \bibinfo {author} {\bibfnamefont {M.}~\bibnamefont {Domingo}},
  \bibinfo {author} {\bibfnamefont {C.}~\bibnamefont {Millán}}, \bibinfo
  {author} {\bibfnamefont {R.}~\bibnamefont {Luna}}, \bibinfo {author}
  {\bibfnamefont {R.}~\bibnamefont {Vilaplana}},\ and\ \bibinfo {author}
  {\bibfnamefont {C.}~\bibnamefont {Santonja}},\ }\bibfield  {title} {\enquote
  {\bibinfo {title} {Density of {CH4}, {N2} and {CO2} ices at different
  temperatures of deposition},}\ }\href
  {https://doi.org/10.1016/j.pss.2008.07.015} {\bibfield  {journal} {\bibinfo
  {journal} {Planetary and Space Science}\ }\textbf {\bibinfo {volume} {56}},\
  \bibinfo {pages} {1748--1752} (\bibinfo {year} {2008})}\BibitemShut {NoStop}%
\bibitem [{\citenamefont {So/rensen}\ and\ \citenamefont
  {Schou}(1978)}]{so/rensen1978}%
  \BibitemOpen
  \bibfield  {author} {\bibinfo {author} {\bibfnamefont {H.}~\bibnamefont
  {So/rensen}}\ and\ \bibinfo {author} {\bibfnamefont {J.}~\bibnamefont
  {Schou}},\ }\bibfield  {title} {\enquote {\bibinfo {title} {Interaction
  between solid nitrogen and 1–3‐{keV} electrons},}\ }\href
  {https://doi.org/10.1063/1.324433} {\bibfield  {journal} {\bibinfo  {journal}
  {Journal of Applied Physics}\ }\textbf {\bibinfo {volume} {49}},\ \bibinfo
  {pages} {5311--5318} (\bibinfo {year} {1978})}\BibitemShut {NoStop}%
\bibitem [{\citenamefont {Hudel}, \citenamefont {Steinacker},\ and\
  \citenamefont {Feulner}(1992)}]{hudel1992}%
  \BibitemOpen
  \bibfield  {author} {\bibinfo {author} {\bibfnamefont {E.}~\bibnamefont
  {Hudel}}, \bibinfo {author} {\bibfnamefont {E.}~\bibnamefont {Steinacker}},\
  and\ \bibinfo {author} {\bibfnamefont {P.}~\bibnamefont {Feulner}},\
  }\bibfield  {title} {\enquote {\bibinfo {title} {Kinetic energy distributions
  of particles desorbed from solid {N2}, {O2}, and {NO} by electron impact},}\
  }\href {https://doi.org/10.1016/0039-6028(92)90077-J} {\bibfield  {journal}
  {\bibinfo  {journal} {Surface Science}\ }\textbf {\bibinfo {volume} {273}},\
  \bibinfo {pages} {405--412} (\bibinfo {year} {1992})}\BibitemShut {NoStop}%
\bibitem [{\citenamefont {Savchenko}\ \emph
  {et~al.}(2017{\natexlab{a}})\citenamefont {Savchenko}, \citenamefont
  {Khyzhniy}, \citenamefont {Uyutnov}, \citenamefont {Bludov}, \citenamefont
  {Barabashov}, \citenamefont {Gumenchuk},\ and\ \citenamefont
  {Bondybey}}]{savchenko2017a}%
  \BibitemOpen
  \bibfield  {author} {\bibinfo {author} {\bibfnamefont {E.}~\bibnamefont
  {Savchenko}}, \bibinfo {author} {\bibfnamefont {I.}~\bibnamefont {Khyzhniy}},
  \bibinfo {author} {\bibfnamefont {S.}~\bibnamefont {Uyutnov}}, \bibinfo
  {author} {\bibfnamefont {M.}~\bibnamefont {Bludov}}, \bibinfo {author}
  {\bibfnamefont {A.}~\bibnamefont {Barabashov}}, \bibinfo {author}
  {\bibfnamefont {G.}~\bibnamefont {Gumenchuk}},\ and\ \bibinfo {author}
  {\bibfnamefont {V.}~\bibnamefont {Bondybey}},\ }\bibfield  {title} {\enquote
  {\bibinfo {title} {Radiation effects in nitrogen and methane “ices”},}\
  }\href {https://doi.org/10.1016/j.nimb.2017.10.014} {\bibfield  {journal}
  {\bibinfo  {journal} {Nuclear Instruments and Methods in Physics Research
  Section B: Beam Interactions with Materials and Atoms}\ } (\bibinfo {year}
  {2017}{\natexlab{a}}),\ 10.1016/j.nimb.2017.10.014}\BibitemShut {NoStop}%
\bibitem [{\citenamefont {Savchenko}\ \emph
  {et~al.}(2017{\natexlab{b}})\citenamefont {Savchenko}, \citenamefont
  {Khyzhniy}, \citenamefont {Uyutnov}, \citenamefont {Bludov}, \citenamefont
  {Barabashov}, \citenamefont {Gumenchuk},\ and\ \citenamefont
  {Bondybey}}]{savchenko2017}%
  \BibitemOpen
  \bibfield  {author} {\bibinfo {author} {\bibfnamefont {E.}~\bibnamefont
  {Savchenko}}, \bibinfo {author} {\bibfnamefont {I.}~\bibnamefont {Khyzhniy}},
  \bibinfo {author} {\bibfnamefont {S.}~\bibnamefont {Uyutnov}}, \bibinfo
  {author} {\bibfnamefont {M.}~\bibnamefont {Bludov}}, \bibinfo {author}
  {\bibfnamefont {A.}~\bibnamefont {Barabashov}}, \bibinfo {author}
  {\bibfnamefont {G.}~\bibnamefont {Gumenchuk}},\ and\ \bibinfo {author}
  {\bibfnamefont {V.}~\bibnamefont {Bondybey}},\ }\bibfield  {title} {\enquote
  {\bibinfo {title} {Excitation-induced processes in model molecular solid –
  {N} 2},}\ }\href {https://doi.org/10.1016/j.jlumin.2016.12.055} {\bibfield
  {journal} {\bibinfo  {journal} {Journal of Luminescence}\ }\textbf {\bibinfo
  {volume} {191}},\ \bibinfo {pages} {73--77} (\bibinfo {year}
  {2017}{\natexlab{b}})}\BibitemShut {NoStop}%
\bibitem [{\citenamefont {Jamieson}, \citenamefont {Mebel},\ and\ \citenamefont
  {Kaiser}(2006)}]{jamieson2006}%
  \BibitemOpen
  \bibfield  {author} {\bibinfo {author} {\bibfnamefont {C.~S.}\ \bibnamefont
  {Jamieson}}, \bibinfo {author} {\bibfnamefont {A.~M.}\ \bibnamefont
  {Mebel}},\ and\ \bibinfo {author} {\bibfnamefont {R.~I.}\ \bibnamefont
  {Kaiser}},\ }\bibfield  {title} {\enquote {\bibinfo {title} {Understanding
  the {Kinetics} and {Dynamics} of {Radiation}‐induced {Reaction} {Pathways}
  in {Carbon} {Monoxide} {Ice} at 10 {K}},}\ }\href
  {https://doi.org/10.1086/499245} {\bibfield  {journal} {\bibinfo  {journal}
  {The Astrophysical Journal Supplement Series}\ }\textbf {\bibinfo {volume}
  {163}},\ \bibinfo {pages} {184--206} (\bibinfo {year} {2006})}\BibitemShut
  {NoStop}%
\bibitem [{\citenamefont {Ellegaard}\ \emph {et~al.}(1988)\citenamefont
  {Ellegaard}, \citenamefont {Pedrys}, \citenamefont {Schou}, \citenamefont
  {Sørensen},\ and\ \citenamefont {Børgesen}}]{ellegaard1988}%
  \BibitemOpen
  \bibfield  {author} {\bibinfo {author} {\bibfnamefont {O.}~\bibnamefont
  {Ellegaard}}, \bibinfo {author} {\bibfnamefont {R.}~\bibnamefont {Pedrys}},
  \bibinfo {author} {\bibfnamefont {J.}~\bibnamefont {Schou}}, \bibinfo
  {author} {\bibfnamefont {H.}~\bibnamefont {Sørensen}},\ and\ \bibinfo
  {author} {\bibfnamefont {P.}~\bibnamefont {Børgesen}},\ }\bibfield  {title}
  {\enquote {\bibinfo {title} {Sputtering of solid argon by {keV} electrons},}\
  }\href {https://doi.org/10.1007/BF01210351} {\bibfield  {journal} {\bibinfo
  {journal} {Applied Physics A}\ }\textbf {\bibinfo {volume} {46}},\ \bibinfo
  {pages} {305--312} (\bibinfo {year} {1988})}\BibitemShut {NoStop}%
\bibitem [{\citenamefont {He}\ and\ \citenamefont {Vidali}(2018)}]{he2018}%
  \BibitemOpen
  \bibfield  {author} {\bibinfo {author} {\bibfnamefont {J.}~\bibnamefont
  {He}}\ and\ \bibinfo {author} {\bibfnamefont {G.}~\bibnamefont {Vidali}},\
  }\bibfield  {title} {\enquote {\bibinfo {title} {Characterization of thin
  film {CO2} ice through the infrared nu1+nu3 combination mode},}\ }\href
  {https://doi.org/10.1093/mnras/stx2412} {\bibfield  {journal} {\bibinfo
  {journal} {Monthly Notices of the Royal Astronomical Society}\ }\textbf
  {\bibinfo {volume} {473}},\ \bibinfo {pages} {860--866} (\bibinfo {year}
  {2018})}\BibitemShut {NoStop}%
\bibitem [{\citenamefont {Schulze}\ and\ \citenamefont
  {Abe}(1980)}]{schulze1980}%
  \BibitemOpen
  \bibfield  {author} {\bibinfo {author} {\bibfnamefont {W.}~\bibnamefont
  {Schulze}}\ and\ \bibinfo {author} {\bibfnamefont {H.}~\bibnamefont {Abe}},\
  }\bibfield  {title} {\enquote {\bibinfo {title} {Density, refractive index
  and sorption capacity of solid {CO2} layers},}\ }\href
  {https://doi.org/10.1016/0301-0104(80)85240-2} {\bibfield  {journal}
  {\bibinfo  {journal} {Chemical Physics}\ }\textbf {\bibinfo {volume} {52}},\
  \bibinfo {pages} {381--388} (\bibinfo {year} {1980})}\BibitemShut {NoStop}%
\bibitem [{\citenamefont {Fillion}\ \emph {et~al.}(2014)\citenamefont
  {Fillion}, \citenamefont {Fayolle}, \citenamefont {Michaut}, \citenamefont
  {Doronin}, \citenamefont {Philippe}, \citenamefont {Rakovsky}, \citenamefont
  {Romanzin}, \citenamefont {Champion}, \citenamefont {Öberg}, \citenamefont
  {Linnartz},\ and\ \citenamefont {Bertin}}]{fillion2014}%
  \BibitemOpen
  \bibfield  {author} {\bibinfo {author} {\bibfnamefont {J.-H.}\ \bibnamefont
  {Fillion}}, \bibinfo {author} {\bibfnamefont {E.~C.}\ \bibnamefont
  {Fayolle}}, \bibinfo {author} {\bibfnamefont {X.}~\bibnamefont {Michaut}},
  \bibinfo {author} {\bibfnamefont {M.}~\bibnamefont {Doronin}}, \bibinfo
  {author} {\bibfnamefont {L.}~\bibnamefont {Philippe}}, \bibinfo {author}
  {\bibfnamefont {J.}~\bibnamefont {Rakovsky}}, \bibinfo {author}
  {\bibfnamefont {C.}~\bibnamefont {Romanzin}}, \bibinfo {author}
  {\bibfnamefont {N.}~\bibnamefont {Champion}}, \bibinfo {author}
  {\bibfnamefont {K.~I.}\ \bibnamefont {Öberg}}, \bibinfo {author}
  {\bibfnamefont {H.}~\bibnamefont {Linnartz}},\ and\ \bibinfo {author}
  {\bibfnamefont {M.}~\bibnamefont {Bertin}},\ }\bibfield  {title} {\enquote
  {\bibinfo {title} {Wavelength resolved {UV} photodesorption and
  photochemistry of {CO2} ice},}\ }\href {https://doi.org/10.1039/C3FD00129F}
  {\bibfield  {journal} {\bibinfo  {journal} {Faraday Discussions}\ }\textbf
  {\bibinfo {volume} {168}},\ \bibinfo {pages} {533} (\bibinfo {year}
  {2014})}\BibitemShut {NoStop}%
\bibitem [{\citenamefont {Sie}\ \emph {et~al.}(2019)\citenamefont {Sie},
  \citenamefont {Caro}, \citenamefont {Huang}, \citenamefont
  {Martín-Doménech}, \citenamefont {Fuente},\ and\ \citenamefont
  {Chen}}]{sie2019}%
  \BibitemOpen
  \bibfield  {author} {\bibinfo {author} {\bibfnamefont {N.-E.}\ \bibnamefont
  {Sie}}, \bibinfo {author} {\bibfnamefont {G.~M.~M.}\ \bibnamefont {Caro}},
  \bibinfo {author} {\bibfnamefont {Z.-H.}\ \bibnamefont {Huang}}, \bibinfo
  {author} {\bibfnamefont {R.}~\bibnamefont {Martín-Doménech}}, \bibinfo
  {author} {\bibfnamefont {A.}~\bibnamefont {Fuente}},\ and\ \bibinfo {author}
  {\bibfnamefont {Y.-J.}\ \bibnamefont {Chen}},\ }\bibfield  {title} {\enquote
  {\bibinfo {title} {On the {Photodesorption} of {CO} $_{\textrm{2}}$ {Ice}
  {Analogs}: {The} {Formation} of {Atomic} {C} in the {Ice} and the {Effect} of
  the {VUV} {Emission} {Spectrum}},}\ }\href
  {https://doi.org/10.3847/1538-4357/ab06be} {\bibfield  {journal} {\bibinfo
  {journal} {The Astrophysical Journal}\ }\textbf {\bibinfo {volume} {874}},\
  \bibinfo {pages} {35} (\bibinfo {year} {2019})}\BibitemShut {NoStop}%
\bibitem [{\citenamefont {Hilleret}\ \emph {et~al.}(2000)\citenamefont
  {Hilleret}, \citenamefont {Baglin}, \citenamefont {Henrist}, \citenamefont
  {Mercier},\ and\ \citenamefont {Scheuerlein}}]{hilleret2000}%
  \BibitemOpen
  \bibfield  {author} {\bibinfo {author} {\bibfnamefont {N.}~\bibnamefont
  {Hilleret}}, \bibinfo {author} {\bibfnamefont {V.}~\bibnamefont {Baglin}},
  \bibinfo {author} {\bibfnamefont {B.}~\bibnamefont {Henrist}}, \bibinfo
  {author} {\bibfnamefont {E.}~\bibnamefont {Mercier}},\ and\ \bibinfo {author}
  {\bibfnamefont {C.}~\bibnamefont {Scheuerlein}},\ }\bibfield  {title}
  {\enquote {\bibinfo {title} {Ingredients for the {Understanding} and the
  {Simulation} of {Multipacting}},}\ }in\ \href@noop {} {\emph {\bibinfo
  {booktitle} {10th {Workshop} on {LEP}-{SPS} {Performance}}}}\ (\bibinfo
  {year} {2000})\ pp.\ \bibinfo {pages} {130--135}\BibitemShut {NoStop}%
\bibitem [{\citenamefont {Dupuy}(2019)}]{dupuy2019e}%
  \BibitemOpen
  \bibfield  {author} {\bibinfo {author} {\bibfnamefont {R.}~\bibnamefont
  {Dupuy}},\ }\emph {\bibinfo {title} {Photon and electron induced desorption
  from molecular ices}},\ \href@noop {} {Ph.D. thesis},\ \bibinfo  {school}
  {Sorbonne Université} (\bibinfo {year} {2019})\BibitemShut {NoStop}%
\bibitem [{\citenamefont {Kimmel}\ \emph {et~al.}(1994)\citenamefont {Kimmel},
  \citenamefont {Orlando}, \citenamefont {Vézina},\ and\ \citenamefont
  {Sanche}}]{kimmel1994}%
  \BibitemOpen
  \bibfield  {author} {\bibinfo {author} {\bibfnamefont {G.~A.}\ \bibnamefont
  {Kimmel}}, \bibinfo {author} {\bibfnamefont {T.~M.}\ \bibnamefont {Orlando}},
  \bibinfo {author} {\bibfnamefont {C.}~\bibnamefont {Vézina}},\ and\ \bibinfo
  {author} {\bibfnamefont {L.}~\bibnamefont {Sanche}},\ }\bibfield  {title}
  {\enquote {\bibinfo {title} {Low‐energy electron‐stimulated production of
  molecular hydrogen from amorphous water ice},}\ }\href
  {https://doi.org/10.1063/1.468430} {\bibfield  {journal} {\bibinfo  {journal}
  {The Journal of Chemical Physics}\ }\textbf {\bibinfo {volume} {101}},\
  \bibinfo {pages} {3282--3286} (\bibinfo {year} {1994})}\BibitemShut {NoStop}%
\bibitem [{\citenamefont {Kimmel}\ and\ \citenamefont
  {Orlando}(1995)}]{kimmel1995}%
  \BibitemOpen
  \bibfield  {author} {\bibinfo {author} {\bibfnamefont {G.~A.}\ \bibnamefont
  {Kimmel}}\ and\ \bibinfo {author} {\bibfnamefont {T.~M.}\ \bibnamefont
  {Orlando}},\ }\bibfield  {title} {\enquote {\bibinfo {title} {Low-{Energy}
  (5–120 {eV}) {Electron}-{Stimulated} {Dissociation} of {Amorphous} {D2O}
  {Ice}: {D}({2S}), {O}({3P2}, 1, 0), and {O}({1D2}) {Yields} and {Velocity}
  {Distributions}},}\ }\href {https://doi.org/10.1103/PhysRevLett.75.2606}
  {\bibfield  {journal} {\bibinfo  {journal} {Physical Review Letters}\
  }\textbf {\bibinfo {volume} {75}},\ \bibinfo {pages} {2606--2609} (\bibinfo
  {year} {1995})}\BibitemShut {NoStop}%
\bibitem [{\citenamefont {Petrik}\ and\ \citenamefont
  {Kimmel}(2004)}]{petrik2004}%
  \BibitemOpen
  \bibfield  {author} {\bibinfo {author} {\bibfnamefont {N.~G.}\ \bibnamefont
  {Petrik}}\ and\ \bibinfo {author} {\bibfnamefont {G.~A.}\ \bibnamefont
  {Kimmel}},\ }\bibfield  {title} {\enquote {\bibinfo {title}
  {Electron-stimulated production of molecular hydrogen at the interfaces of
  amorphous solid water films on {Pt}(111)},}\ }\href
  {https://doi.org/10.1063/1.1773152} {\bibfield  {journal} {\bibinfo
  {journal} {The Journal of Chemical Physics}\ }\textbf {\bibinfo {volume}
  {121}},\ \bibinfo {pages} {3736--3744} (\bibinfo {year} {2004})}\BibitemShut
  {NoStop}%
\bibitem [{\citenamefont {Petrik}\ and\ \citenamefont
  {Kimmel}(2005)}]{petrik2005}%
  \BibitemOpen
  \bibfield  {author} {\bibinfo {author} {\bibfnamefont {N.~G.}\ \bibnamefont
  {Petrik}}\ and\ \bibinfo {author} {\bibfnamefont {G.~A.}\ \bibnamefont
  {Kimmel}},\ }\bibfield  {title} {\enquote {\bibinfo {title}
  {Electron-stimulated sputtering of thin amorphous solid water films on
  {Pt}(111)},}\ }\href {https://doi.org/10.1063/1.1943388} {\bibfield
  {journal} {\bibinfo  {journal} {The Journal of Chemical Physics}\ }\textbf
  {\bibinfo {volume} {123}},\ \bibinfo {pages} {054702} (\bibinfo {year}
  {2005})}\BibitemShut {NoStop}%
\bibitem [{\citenamefont {Petrik}, \citenamefont {Kavetsky},\ and\
  \citenamefont {Kimmel}(2006)}]{petrik2006a}%
  \BibitemOpen
  \bibfield  {author} {\bibinfo {author} {\bibfnamefont {N.~G.}\ \bibnamefont
  {Petrik}}, \bibinfo {author} {\bibfnamefont {A.~G.}\ \bibnamefont
  {Kavetsky}},\ and\ \bibinfo {author} {\bibfnamefont {G.~A.}\ \bibnamefont
  {Kimmel}},\ }\bibfield  {title} {\enquote {\bibinfo {title}
  {Electron-stimulated production of molecular oxygen in amorphous solid water
  on {Pt}(111): {Precursor} transport through the hydrogen bonding network},}\
  }\href {https://doi.org/10.1063/1.2345367} {\bibfield  {journal} {\bibinfo
  {journal} {The Journal of Chemical Physics}\ }\textbf {\bibinfo {volume}
  {125}},\ \bibinfo {pages} {124702} (\bibinfo {year} {2006})}\BibitemShut
  {NoStop}%
\bibitem [{\citenamefont {Marchione}\ and\ \citenamefont
  {McCoustra}(2016)}]{marchione2016a}%
  \BibitemOpen
  \bibfield  {author} {\bibinfo {author} {\bibfnamefont {D.}~\bibnamefont
  {Marchione}}\ and\ \bibinfo {author} {\bibfnamefont {M.~R.~S.}\ \bibnamefont
  {McCoustra}},\ }\bibfield  {title} {\enquote {\bibinfo {title} {Electrons,
  excitons and hydrogen bonding: electron-promoted desorption from molecular
  ice surfaces},}\ }\href {https://doi.org/10.1039/C6CP05814K} {\bibfield
  {journal} {\bibinfo  {journal} {Phys. Chem. Chem. Phys.}\ }\textbf {\bibinfo
  {volume} {18}},\ \bibinfo {pages} {29747--29755} (\bibinfo {year}
  {2016})}\BibitemShut {NoStop}%
\bibitem [{\citenamefont {Abdulgalil}\ \emph {et~al.}(2017)\citenamefont
  {Abdulgalil}, \citenamefont {Rosu-Finsen}, \citenamefont {Marchione},
  \citenamefont {Thrower}, \citenamefont {Collings},\ and\ \citenamefont
  {McCoustra}}]{abdulgalil2017}%
  \BibitemOpen
  \bibfield  {author} {\bibinfo {author} {\bibfnamefont {A.~G.~M.}\
  \bibnamefont {Abdulgalil}}, \bibinfo {author} {\bibfnamefont
  {A.}~\bibnamefont {Rosu-Finsen}}, \bibinfo {author} {\bibfnamefont
  {D.}~\bibnamefont {Marchione}}, \bibinfo {author} {\bibfnamefont {J.~D.}\
  \bibnamefont {Thrower}}, \bibinfo {author} {\bibfnamefont {M.~P.}\
  \bibnamefont {Collings}},\ and\ \bibinfo {author} {\bibfnamefont {M.~R.~S.}\
  \bibnamefont {McCoustra}},\ }\bibfield  {title} {\enquote {\bibinfo {title}
  {Electron-{Promoted} {Desorption} from {Water} {Ice} {Surfaces}: {Neutral}
  {Gas}-{Phase} {Products}},}\ }\href
  {https://doi.org/10.1021/acsearthspacechem.7b00028} {\bibfield  {journal}
  {\bibinfo  {journal} {ACS Earth and Space Chemistry}\ }\textbf {\bibinfo
  {volume} {1}},\ \bibinfo {pages} {209--215} (\bibinfo {year}
  {2017})}\BibitemShut {NoStop}%
\bibitem [{\citenamefont {Schou}(1987)}]{schou1987}%
  \BibitemOpen
  \bibfield  {author} {\bibinfo {author} {\bibfnamefont {J.}~\bibnamefont
  {Schou}},\ }\bibfield  {title} {\enquote {\bibinfo {title} {Sputtering of
  frozen gases},}\ }\href {https://doi.org/10.1016/0168-583X(87)90020-6}
  {\bibfield  {journal} {\bibinfo  {journal} {Nuclear Instruments and Methods
  in Physics Research Section B: Beam Interactions with Materials and Atoms}\
  }\textbf {\bibinfo {volume} {27}},\ \bibinfo {pages} {188--200} (\bibinfo
  {year} {1987})}\BibitemShut {NoStop}%
\bibitem [{\citenamefont {Famá}\ \emph {et~al.}(2007)\citenamefont {Famá},
  \citenamefont {Teolis}, \citenamefont {Bahr},\ and\ \citenamefont
  {Baragiola}}]{fama2007}%
  \BibitemOpen
  \bibfield  {author} {\bibinfo {author} {\bibfnamefont {M.}~\bibnamefont
  {Famá}}, \bibinfo {author} {\bibfnamefont {B.~D.}\ \bibnamefont {Teolis}},
  \bibinfo {author} {\bibfnamefont {D.~A.}\ \bibnamefont {Bahr}},\ and\
  \bibinfo {author} {\bibfnamefont {R.~A.}\ \bibnamefont {Baragiola}},\
  }\bibfield  {title} {\enquote {\bibinfo {title} {Role of electron capture in
  ion-induced electronic sputtering of insulators},}\ }\href
  {https://doi.org/10.1103/PhysRevB.75.100101} {\bibfield  {journal} {\bibinfo
  {journal} {Phys. Rev. B}\ }\textbf {\bibinfo {volume} {75}},\ \bibinfo
  {pages} {100101} (\bibinfo {year} {2007})}\BibitemShut {NoStop}%
\bibitem [{\citenamefont {Chrisey}, \citenamefont {Brown},\ and\ \citenamefont
  {Boring}(1990)}]{chrisey1990}%
  \BibitemOpen
  \bibfield  {author} {\bibinfo {author} {\bibfnamefont {D.~B.}\ \bibnamefont
  {Chrisey}}, \bibinfo {author} {\bibfnamefont {W.~L.}\ \bibnamefont {Brown}},\
  and\ \bibinfo {author} {\bibfnamefont {J.~W.}\ \bibnamefont {Boring}},\
  }\bibfield  {title} {\enquote {\bibinfo {title} {Electronic excitation of
  condensed {CO}: sputtering and chemical change},}\ }\href
  {https://doi.org/10.1016/0039-6028(90)90431-7} {\bibfield  {journal}
  {\bibinfo  {journal} {Surface science}\ }\textbf {\bibinfo {volume} {225}},\
  \bibinfo {pages} {130--142} (\bibinfo {year} {1990})}\BibitemShut {NoStop}%
\bibitem [{\citenamefont {Seperuelo~Duarte}\ \emph {et~al.}(2010)\citenamefont
  {Seperuelo~Duarte}, \citenamefont {Domaracka}, \citenamefont {Boduch},
  \citenamefont {Rothard}, \citenamefont {Dartois},\ and\ \citenamefont
  {da~Silveira}}]{seperueloduarte2010}%
  \BibitemOpen
  \bibfield  {author} {\bibinfo {author} {\bibfnamefont {E.}~\bibnamefont
  {Seperuelo~Duarte}}, \bibinfo {author} {\bibfnamefont {A.}~\bibnamefont
  {Domaracka}}, \bibinfo {author} {\bibfnamefont {P.}~\bibnamefont {Boduch}},
  \bibinfo {author} {\bibfnamefont {H.}~\bibnamefont {Rothard}}, \bibinfo
  {author} {\bibfnamefont {E.}~\bibnamefont {Dartois}},\ and\ \bibinfo {author}
  {\bibfnamefont {E.~F.}\ \bibnamefont {da~Silveira}},\ }\bibfield  {title}
  {\enquote {\bibinfo {title} {Laboratory simulation of heavy-ion cosmic-ray
  interaction with condensed {CO}},}\ }\href
  {https://doi.org/10.1051/0004-6361/200912899} {\bibfield  {journal} {\bibinfo
   {journal} {Astronomy and Astrophysics}\ }\textbf {\bibinfo {volume} {512}},\
  \bibinfo {pages} {A71} (\bibinfo {year} {2010})}\BibitemShut {NoStop}%
\bibitem [{\citenamefont {Brown}\ \emph {et~al.}(1982)\citenamefont {Brown},
  \citenamefont {Augustyniak}, \citenamefont {Simmons}, \citenamefont
  {Marcantonio}, \citenamefont {Lanzerotti}, \citenamefont {Johnson},
  \citenamefont {Boring}, \citenamefont {Reimann}, \citenamefont {Foti},\ and\
  \citenamefont {Pirronello}}]{brown1982}%
  \BibitemOpen
  \bibfield  {author} {\bibinfo {author} {\bibfnamefont {W.}~\bibnamefont
  {Brown}}, \bibinfo {author} {\bibfnamefont {W.}~\bibnamefont {Augustyniak}},
  \bibinfo {author} {\bibfnamefont {E.}~\bibnamefont {Simmons}}, \bibinfo
  {author} {\bibfnamefont {K.}~\bibnamefont {Marcantonio}}, \bibinfo {author}
  {\bibfnamefont {L.}~\bibnamefont {Lanzerotti}}, \bibinfo {author}
  {\bibfnamefont {R.}~\bibnamefont {Johnson}}, \bibinfo {author} {\bibfnamefont
  {J.}~\bibnamefont {Boring}}, \bibinfo {author} {\bibfnamefont
  {C.}~\bibnamefont {Reimann}}, \bibinfo {author} {\bibfnamefont
  {G.}~\bibnamefont {Foti}},\ and\ \bibinfo {author} {\bibfnamefont
  {V.}~\bibnamefont {Pirronello}},\ }\bibfield  {title} {\enquote {\bibinfo
  {title} {Erosion and molecule formation in condensed gas films by electronic
  energy loss of fast ions},}\ }\href
  {https://doi.org/10.1016/0167-5087(82)90043-6} {\bibfield  {journal}
  {\bibinfo  {journal} {Nuclear Instruments and Methods in Physics Research}\
  }\textbf {\bibinfo {volume} {198}},\ \bibinfo {pages} {1--8} (\bibinfo {year}
  {1982})}\BibitemShut {NoStop}%
\bibitem [{\citenamefont {Seperuelo~Duarte}\ \emph {et~al.}(2009)\citenamefont
  {Seperuelo~Duarte}, \citenamefont {Boduch}, \citenamefont {Rothard},
  \citenamefont {Been}, \citenamefont {Dartois}, \citenamefont {Farenzena},\
  and\ \citenamefont {da~Silveira}}]{seperueloduarte2009}%
  \BibitemOpen
  \bibfield  {author} {\bibinfo {author} {\bibfnamefont {E.}~\bibnamefont
  {Seperuelo~Duarte}}, \bibinfo {author} {\bibfnamefont {P.}~\bibnamefont
  {Boduch}}, \bibinfo {author} {\bibfnamefont {H.}~\bibnamefont {Rothard}},
  \bibinfo {author} {\bibfnamefont {T.}~\bibnamefont {Been}}, \bibinfo {author}
  {\bibfnamefont {E.}~\bibnamefont {Dartois}}, \bibinfo {author} {\bibfnamefont
  {L.~S.}\ \bibnamefont {Farenzena}},\ and\ \bibinfo {author} {\bibfnamefont
  {E.~F.}\ \bibnamefont {da~Silveira}},\ }\bibfield  {title} {\enquote
  {\bibinfo {title} {Heavy ion irradiation of condensed
  {CO}\${\textbackslash}mathsf\{\_\{2\}\}\$: sputtering and molecule
  formation},}\ }\href {https://doi.org/10.1051/0004-6361/200811359} {\bibfield
   {journal} {\bibinfo  {journal} {A\&A}\ }\textbf {\bibinfo {volume} {502}},\
  \bibinfo {pages} {599--603} (\bibinfo {year} {2009})}\BibitemShut {NoStop}%
\bibitem [{\citenamefont {Raut}\ and\ \citenamefont
  {Baragiola}(2013)}]{raut2013}%
  \BibitemOpen
  \bibfield  {author} {\bibinfo {author} {\bibfnamefont {U.}~\bibnamefont
  {Raut}}\ and\ \bibinfo {author} {\bibfnamefont {R.~A.}\ \bibnamefont
  {Baragiola}},\ }\bibfield  {title} {\enquote {\bibinfo {title} {Sputtering
  and {Molecular} {Synthesis} {Induced} by 100 {Kev} {Protons} in {Condensed}
  {CO}$_{\textrm{2}}$ and {Relevance} to the {Outer} {Solar} {System}},}\
  }\href {https://doi.org/10.1088/0004-637X/772/1/53} {\bibfield  {journal}
  {\bibinfo  {journal} {ApJ}\ }\textbf {\bibinfo {volume} {772}},\ \bibinfo
  {pages} {53} (\bibinfo {year} {2013})}\BibitemShut {NoStop}%
\bibitem [{\citenamefont {Mejía}\ \emph {et~al.}(2015)\citenamefont {Mejía},
  \citenamefont {Bender}, \citenamefont {Severin}, \citenamefont {Trautmann},
  \citenamefont {Boduch}, \citenamefont {Bordalo}, \citenamefont {Domaracka},
  \citenamefont {Lv}, \citenamefont {Martinez},\ and\ \citenamefont
  {Rothard}}]{mejia2015}%
  \BibitemOpen
  \bibfield  {author} {\bibinfo {author} {\bibfnamefont {C.}~\bibnamefont
  {Mejía}}, \bibinfo {author} {\bibfnamefont {M.}~\bibnamefont {Bender}},
  \bibinfo {author} {\bibfnamefont {D.}~\bibnamefont {Severin}}, \bibinfo
  {author} {\bibfnamefont {C.}~\bibnamefont {Trautmann}}, \bibinfo {author}
  {\bibfnamefont {P.}~\bibnamefont {Boduch}}, \bibinfo {author} {\bibfnamefont
  {V.}~\bibnamefont {Bordalo}}, \bibinfo {author} {\bibfnamefont
  {A.}~\bibnamefont {Domaracka}}, \bibinfo {author} {\bibfnamefont
  {X.}~\bibnamefont {Lv}}, \bibinfo {author} {\bibfnamefont {R.}~\bibnamefont
  {Martinez}},\ and\ \bibinfo {author} {\bibfnamefont {H.}~\bibnamefont
  {Rothard}},\ }\bibfield  {title} {\enquote {\bibinfo {title} {Radiolysis and
  sputtering of carbon dioxide ice induced by swift {Ti}, {Ni}, and {Xe}
  ions},}\ }\href {https://doi.org/10.1016/j.nimb.2015.09.039} {\bibfield
  {journal} {\bibinfo  {journal} {Nuclear Instruments and Methods in Physics
  Research Section B: Beam Interactions with Materials and Atoms}\ }\textbf
  {\bibinfo {volume} {365}},\ \bibinfo {pages} {477--481} (\bibinfo {year}
  {2015})}\BibitemShut {NoStop}%
\bibitem [{\citenamefont {Reimann}, \citenamefont {Brown},\ and\ \citenamefont
  {Johnson}(1988)}]{reimann1988}%
  \BibitemOpen
  \bibfield  {author} {\bibinfo {author} {\bibfnamefont {C.~T.}\ \bibnamefont
  {Reimann}}, \bibinfo {author} {\bibfnamefont {W.~L.}\ \bibnamefont {Brown}},\
  and\ \bibinfo {author} {\bibfnamefont {R.~E.}\ \bibnamefont {Johnson}},\
  }\bibfield  {title} {\enquote {\bibinfo {title} {Electronically stimulated
  sputtering and luminescence from solid argon},}\ }\href
  {https://doi.org/10.1103/PhysRevB.37.1455} {\bibfield  {journal} {\bibinfo
  {journal} {Physical Review B}\ }\textbf {\bibinfo {volume} {37}},\ \bibinfo
  {pages} {1455--1473} (\bibinfo {year} {1988})}\BibitemShut {NoStop}%
\bibitem [{\citenamefont {Johnson}, \citenamefont {Pospieszalska},\ and\
  \citenamefont {Brown}(1991)}]{johnson1991}%
  \BibitemOpen
  \bibfield  {author} {\bibinfo {author} {\bibfnamefont {R.~E.}\ \bibnamefont
  {Johnson}}, \bibinfo {author} {\bibfnamefont {M.}~\bibnamefont
  {Pospieszalska}},\ and\ \bibinfo {author} {\bibfnamefont {W.~L.}\
  \bibnamefont {Brown}},\ }\bibfield  {title} {\enquote {\bibinfo {title}
  {Linear-to-quadratic transition in electronically stimulated sputtering of
  solid {N} 2 and {O} 2},}\ }\href {https://doi.org/10.1103/PhysRevB.44.7263}
  {\bibfield  {journal} {\bibinfo  {journal} {Phys. Rev. B}\ }\textbf {\bibinfo
  {volume} {44}},\ \bibinfo {pages} {7263--7272} (\bibinfo {year}
  {1991})}\BibitemShut {NoStop}%
\bibitem [{\citenamefont {Adams}\ and\ \citenamefont
  {Hansma}(1980)}]{adams1980}%
  \BibitemOpen
  \bibfield  {author} {\bibinfo {author} {\bibfnamefont {A.}~\bibnamefont
  {Adams}}\ and\ \bibinfo {author} {\bibfnamefont {P.~K.}\ \bibnamefont
  {Hansma}},\ }\bibfield  {title} {\enquote {\bibinfo {title} {Practical range
  and energy loss of 0.1-3-{keV} electrons in thin films of {N2} , {O2} , {Ar},
  {Kr}, and {Xe}},}\ }\href {https://doi.org/10.1103/PhysRevB.22.4258}
  {\bibfield  {journal} {\bibinfo  {journal} {Physical Review B}\ }\textbf
  {\bibinfo {volume} {22}},\ \bibinfo {pages} {4258--4263} (\bibinfo {year}
  {1980})}\BibitemShut {NoStop}%
\bibitem [{\citenamefont {Gümüş}(2005)}]{gumus2005}%
  \BibitemOpen
  \bibfield  {author} {\bibinfo {author} {\bibfnamefont {H.}~\bibnamefont
  {Gümüş}},\ }\bibfield  {title} {\enquote {\bibinfo {title} {Simple
  stopping power formula for low and intermediate energy electrons},}\ }\href
  {https://doi.org/10.1016/j.radphyschem.2004.03.006} {\bibfield  {journal}
  {\bibinfo  {journal} {Radiation Physics and Chemistry}\ }\textbf {\bibinfo
  {volume} {72}},\ \bibinfo {pages} {7--12} (\bibinfo {year}
  {2005})}\BibitemShut {NoStop}%
\bibitem [{\citenamefont {LaVerne}\ and\ \citenamefont
  {Mozumder}(1983)}]{laverne1983}%
  \BibitemOpen
  \bibfield  {author} {\bibinfo {author} {\bibfnamefont {J.~A.}\ \bibnamefont
  {LaVerne}}\ and\ \bibinfo {author} {\bibfnamefont {A.}~\bibnamefont
  {Mozumder}},\ }\bibfield  {title} {\enquote {\bibinfo {title} {Penetration of
  {Low}-{Energy} {Electrons} in {Water}},}\ }\href
  {https://doi.org/10.2307/3576206} {\bibfield  {journal} {\bibinfo  {journal}
  {Radiation Research}\ }\textbf {\bibinfo {volume} {96}},\ \bibinfo {pages}
  {219} (\bibinfo {year} {1983})}\BibitemShut {NoStop}%
\bibitem [{\citenamefont {Valkealahti}, \citenamefont {Schou},\ and\
  \citenamefont {Nieminen}(1989)}]{valkealahti1989}%
  \BibitemOpen
  \bibfield  {author} {\bibinfo {author} {\bibfnamefont {S.}~\bibnamefont
  {Valkealahti}}, \bibinfo {author} {\bibfnamefont {J.}~\bibnamefont {Schou}},\
  and\ \bibinfo {author} {\bibfnamefont {R.~M.}\ \bibnamefont {Nieminen}},\
  }\bibfield  {title} {\enquote {\bibinfo {title} {Energy deposition of {keV}
  electrons in light elements},}\ }\href {https://doi.org/10.1063/1.342839}
  {\bibfield  {journal} {\bibinfo  {journal} {Journal of Applied Physics}\
  }\textbf {\bibinfo {volume} {65}},\ \bibinfo {pages} {2258--2266} (\bibinfo
  {year} {1989})}\BibitemShut {NoStop}%
\bibitem [{\citenamefont {Bertin}\ \emph {et~al.}(2012)\citenamefont {Bertin},
  \citenamefont {Fayolle}, \citenamefont {Romanzin}, \citenamefont {Öberg},
  \citenamefont {Michaut}, \citenamefont {Moudens}, \citenamefont {Philippe},
  \citenamefont {Jeseck}, \citenamefont {Linnartz},\ and\ \citenamefont
  {Fillion}}]{bertin2012}%
  \BibitemOpen
  \bibfield  {author} {\bibinfo {author} {\bibfnamefont {M.}~\bibnamefont
  {Bertin}}, \bibinfo {author} {\bibfnamefont {E.~C.}\ \bibnamefont {Fayolle}},
  \bibinfo {author} {\bibfnamefont {C.}~\bibnamefont {Romanzin}}, \bibinfo
  {author} {\bibfnamefont {K.~I.}\ \bibnamefont {Öberg}}, \bibinfo {author}
  {\bibfnamefont {X.}~\bibnamefont {Michaut}}, \bibinfo {author} {\bibfnamefont
  {A.}~\bibnamefont {Moudens}}, \bibinfo {author} {\bibfnamefont
  {L.}~\bibnamefont {Philippe}}, \bibinfo {author} {\bibfnamefont
  {P.}~\bibnamefont {Jeseck}}, \bibinfo {author} {\bibfnamefont
  {H.}~\bibnamefont {Linnartz}},\ and\ \bibinfo {author} {\bibfnamefont
  {J.-H.}\ \bibnamefont {Fillion}},\ }\bibfield  {title} {\enquote {\bibinfo
  {title} {{UV} photodesorption of interstellar {CO} ice analogues: from
  subsurface excitation to surface desorption},}\ }\href
  {https://doi.org/10.1039/c2cp41177f} {\bibfield  {journal} {\bibinfo
  {journal} {Physical Chemistry Chemical Physics}\ }\textbf {\bibinfo {volume}
  {14}},\ \bibinfo {pages} {9929} (\bibinfo {year} {2012})}\BibitemShut
  {NoStop}%
\bibitem [{\citenamefont {Zimmerer}(1994)}]{zimmerer1994}%
  \BibitemOpen
  \bibfield  {author} {\bibinfo {author} {\bibfnamefont {G.}~\bibnamefont
  {Zimmerer}},\ }\bibfield  {title} {\enquote {\bibinfo {title} {Electronic
  sputtering from rare-gas solids},}\ }\href
  {https://doi.org/10.1016/0168-583X(94)96295-2} {\bibfield  {journal}
  {\bibinfo  {journal} {Nuclear Instruments and Methods in Physics Research
  Section B: Beam Interactions with Materials and Atoms}\ }\textbf {\bibinfo
  {volume} {91}},\ \bibinfo {pages} {601--613} (\bibinfo {year}
  {1994})}\BibitemShut {NoStop}%
\bibitem [{\citenamefont {Lu}\ \emph {et~al.}(2005)\citenamefont {Lu},
  \citenamefont {Chen}, \citenamefont {Cheng}, \citenamefont {Kuo},\ and\
  \citenamefont {Ogilvie}}]{lu2005}%
  \BibitemOpen
  \bibfield  {author} {\bibinfo {author} {\bibfnamefont {H.-C.}\ \bibnamefont
  {Lu}}, \bibinfo {author} {\bibfnamefont {H.-K.}\ \bibnamefont {Chen}},
  \bibinfo {author} {\bibfnamefont {B.-M.}\ \bibnamefont {Cheng}}, \bibinfo
  {author} {\bibfnamefont {Y.-P.}\ \bibnamefont {Kuo}},\ and\ \bibinfo {author}
  {\bibfnamefont {J.~F.}\ \bibnamefont {Ogilvie}},\ }\bibfield  {title}
  {\enquote {\bibinfo {title} {Spectra in the vacuum ultraviolet region of {CO}
  in gaseous and solid phases and dispersed in solid argon at 10 {K}},}\ }\href
  {https://doi.org/10.1088/0953-4075/38/20/006} {\bibfield  {journal} {\bibinfo
   {journal} {Journal of Physics B: Atomic, Molecular and Optical Physics}\
  }\textbf {\bibinfo {volume} {38}},\ \bibinfo {pages} {3693--3704} (\bibinfo
  {year} {2005})}\BibitemShut {NoStop}%
\bibitem [{\citenamefont {Fugol'}(1978)}]{fugol1978}%
  \BibitemOpen
  \bibfield  {author} {\bibinfo {author} {\bibfnamefont {I.~Y.}\ \bibnamefont
  {Fugol'}},\ }\bibfield  {title} {\enquote {\bibinfo {title} {Excitons in
  rare-gas crystals},}\ }\href {https://doi.org/10.1080/00018737800101344}
  {\bibfield  {journal} {\bibinfo  {journal} {Advances in Physics}\ }\textbf
  {\bibinfo {volume} {27}},\ \bibinfo {pages} {1--87} (\bibinfo {year}
  {1978})}\BibitemShut {NoStop}%
\bibitem [{\citenamefont {Fayolle}\ \emph {et~al.}(2013)\citenamefont
  {Fayolle}, \citenamefont {Bertin}, \citenamefont {Romanzin}, \citenamefont
  {M~Poderoso}, \citenamefont {Philippe}, \citenamefont {Michaut},
  \citenamefont {Jeseck}, \citenamefont {Linnartz}, \citenamefont {Öberg},\
  and\ \citenamefont {Fillion}}]{fayolle2013}%
  \BibitemOpen
  \bibfield  {author} {\bibinfo {author} {\bibfnamefont {E.~C.}\ \bibnamefont
  {Fayolle}}, \bibinfo {author} {\bibfnamefont {M.}~\bibnamefont {Bertin}},
  \bibinfo {author} {\bibfnamefont {C.}~\bibnamefont {Romanzin}}, \bibinfo
  {author} {\bibfnamefont {H.~A.}\ \bibnamefont {M~Poderoso}}, \bibinfo
  {author} {\bibfnamefont {L.}~\bibnamefont {Philippe}}, \bibinfo {author}
  {\bibfnamefont {X.}~\bibnamefont {Michaut}}, \bibinfo {author} {\bibfnamefont
  {P.}~\bibnamefont {Jeseck}}, \bibinfo {author} {\bibfnamefont
  {H.}~\bibnamefont {Linnartz}}, \bibinfo {author} {\bibfnamefont {K.~I.}\
  \bibnamefont {Öberg}},\ and\ \bibinfo {author} {\bibfnamefont {J.-H.}\
  \bibnamefont {Fillion}},\ }\bibfield  {title} {\enquote {\bibinfo {title}
  {Wavelength-dependent {UV} photodesorption of pure {N2} and {O2} ices},}\
  }\href {https://doi.org/10.1051/0004-6361/201321533} {\bibfield  {journal}
  {\bibinfo  {journal} {Astronomy \& Astrophysics}\ }\textbf {\bibinfo {volume}
  {556}},\ \bibinfo {pages} {A122} (\bibinfo {year} {2013})}\BibitemShut
  {NoStop}%
\bibitem [{\citenamefont {Voreades}(1976)}]{voreades1976}%
  \BibitemOpen
  \bibfield  {author} {\bibinfo {author} {\bibfnamefont {D.}~\bibnamefont
  {Voreades}},\ }\bibfield  {title} {\enquote {\bibinfo {title} {Secondary
  electron emission from thin carbon films},}\ }\href
  {https://doi.org/10.1016/0039-6028(76)90320-4} {\bibfield  {journal}
  {\bibinfo  {journal} {Surface Science}\ }\textbf {\bibinfo {volume} {60}},\
  \bibinfo {pages} {325--348} (\bibinfo {year} {1976})}\BibitemShut {NoStop}%
\end{thebibliography}
\end{document}